# The Mathematical Parallels Between Packet Switching and Information Transmission

Tony T. Lee, *Fellow, IEEE*

***Abstract*— All communication networks comprise of transmission systems and switching systems, even though they are usually treated as two separate issues. Communication channels are generally disturbed by noise from various sources. In circuit switched networks, reliable communication requires the error-tolerant transmission of bits over noisy channels. In packet switched networks, however, not only can bits be corrupted with noise, but resources along connection paths are also subject to contention. Thus, quality of service (QoS) is determined by buffer delays and packet losses. The theme of this paper is to show that transmission noise and packet contention actually have similar characteristics and can be tamed by comparable means to achieve reliable communication. The following analogies between switching and transmission are identified.**

**1. Buffering against contention is a process that is similar to the error correction of noise corrupted signals. A signal-to-noise ratio that represents the carried load of packet switches can be deduced from the Boltzmann model of packet distribution.**

**2. When deflection routing is applied to Clos networks the loss probability decreases exponentially, which is similar to the exponential behavior of the error probability of binary symmetric channels with random channel coding. In information theory, this result is stated as the noisy channel coding theorem.**

**3. The similarity between Hall's condition of bipartite matching and expander graph manifests the resemblance between nonblocking route assignments and error-correcting codes. An extension of the Sipser-Speilman decoding algorithm of expander codes to route assignments of Benes networks is given to illustrate their correspondence.**

**4. Scheduling in packet switching serves the same function as noiseless channel coding in digital transmission. The smoothness of scheduling, like source coding, is bounded by entropy inequalities.**

**5. The sampling theorem of bandlimited signals provides the cornerstone of digital communication and signal processing. Recently, the Birkhoff-von Neumann decomposition of traffic matrices has been widely applied to packet switches. With respect to the complexity reduction of packet switching, we show that the decomposition of a doubly stochastic traffic matrix plays a similar role to that of the sampling theorem in digital transmission.**

**We conclude that packet switching systems are governed by mathematical laws that are similar to those of digital transmission systems as envisioned by Shannon in his seminal 1948 paper, *A Mathematical Theory of Communication*.**

## I. Introduction

All communication networks comprise of transmission systems and switching systems, even though they are usually treated as two separate issues. Communication channels are generally disturbed by noise from various sources. In circuit switched networks, resources are dedicated to connections and reliable communication only requires the error-tolerant transmission of bits over noisy channels. In packet switched networks, however, not only can bits be corrupted with noise, but resources along connection paths are also subject to contention. Despite the great achievements of information theory in dealing with transmission noise [1]–[3], there are still many networking problems in modern communication systems that information theory is unable to provide solutions for. A particular problem that has come to the fore is the delay that is induced by contention in packet switched networks, the solution for which lies in extending the theory to go beyond the bit-rate of transmission [4].

Manuscript received xxxx, xxxx. This work was supported by xxxx.

The authors are with the Department of Information Engineering, the Chinese University of Hong Kong, Shatin, N.T., Hong Kong, China.

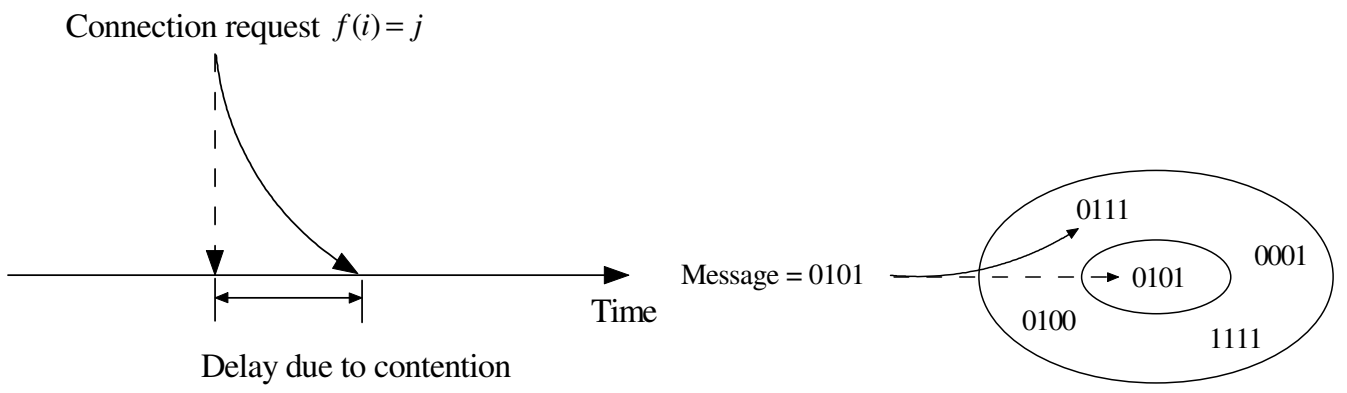

(a) Packet switching with contention (b) Transmission channel with noise

Fig. 1. The parallel characteristics between packet switching and digital transmission

The theme of this paper is to show that transmission noise and packet contention actually have similar mathematical characteristics and can be tamed by comparable means to achieve reliable communication. The source information of the transmission channel is a function of time, and errors are corrected by coding scheme, which is an expansion of signal space. For switching systems, source information is a space function $f(i) = j$, for $i = 1, 2, \cdots, N$ form inputs $V_I$ to outputs $V_o$, which represents a set of connection requests. Packets lost in contention are usually buffered or deflected to stretch out time and defer their requests. As shown in Fig. 1, the two processes are anti-symmetric with respect to time and space and they are both governed by the law of probability.

In transmission systems, the fundamental QoS parameter is bit-rate, or channel capacity, which can be pinpointed by signal-to-noise ratio. In packet switched networks, QoS parameters, buffer delay and packet loss, are all determined by loading. In order to provide a common ground for comparison, the carried load of a packet switch is converted into a pseudo signal-to-noise ratio (PSNR) induced by Boltzmann's model of packet distribution. Based on this correspondence between carried load and PSNR, we show that a packet switched Clos

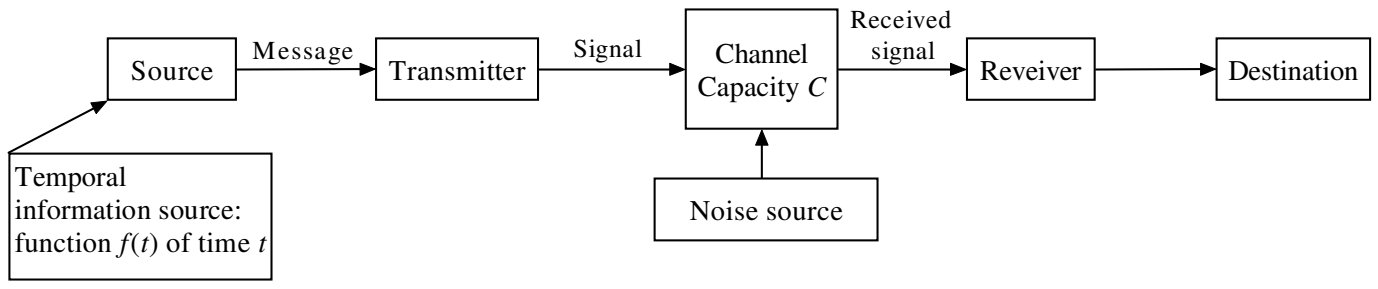

(a) Shannon's general communication system

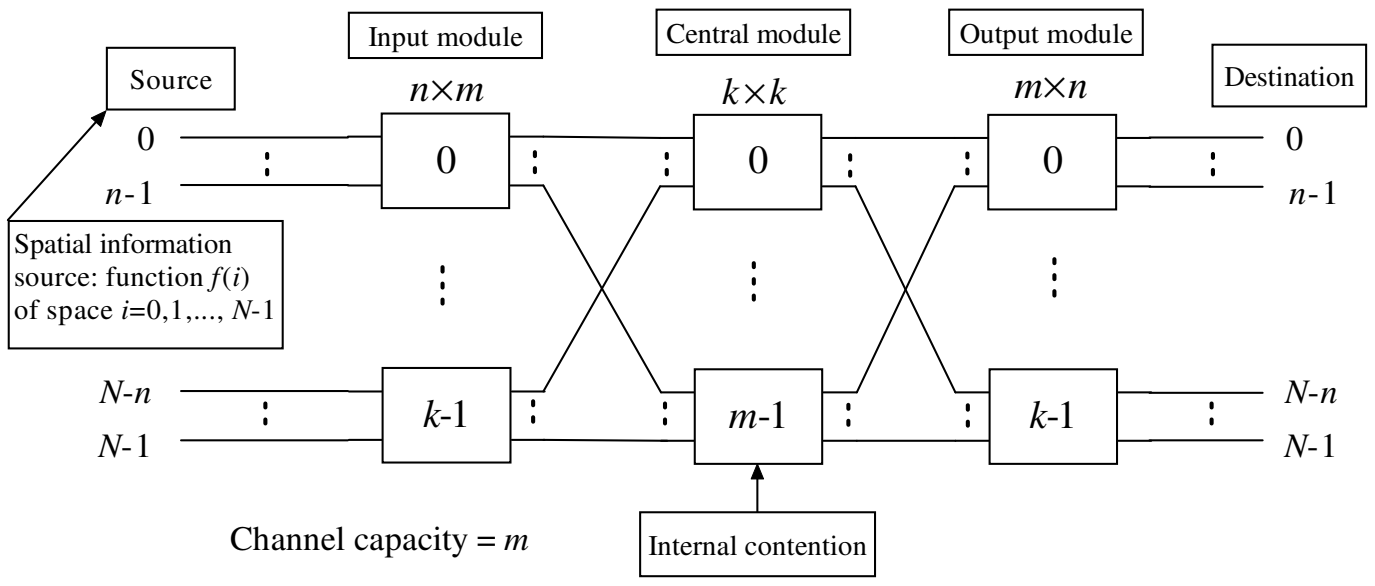

(b) A packet-switched Clos network C$(m, n, k)$

Fig. 2. Comparison of transmission systems and switching systems

network with random routing can be modelled as an abstract channel with additive Gaussian noise.

A schematic diagram of transmission channel and a three-stage Clos network [5] are shown in Fig. 2. The input modules in the first stage and output modules in the third stage of the Clos network correspond to transmitters and receivers, respectively, of the transmission system. The number of central modules is the bandwidth of the Clos network. The disturbances due to internal contentions in the middle stage of a packet switched Clos network mimic the noise of a channel. In this paper, we demonstrate that various routing schemes applied to a packet switched Clos network to cope with contentions are comparable to coding schemes of a transmission channel. The routing schemes of packet switching and their counterparts in digital transmission are listed in Fig.3. Depending on the implementation of routing schemes, and similar to the separation of channel coding and source coding in information theory, the Clos network can either be modelled as a noisy channel with nonblocking routing, analogous to channel coding, or a noiseless channel with route scheduling, analogous to source coding, as explained below.

***The noisy channel model of the Clos network***. We first compare the deflection routing of packet switching with the random coding over noisy channel. Errors introduced by noises in transmission are subdued by redundant bits, while packet contentions can be relieved by redundant links in switching. When deflection routing is applied to Clos networks, the loss probability decreases exponentially with respect to the network length, which is similar to the exponential behavior of the error probability of binary symmetric channels with random channel coding. In information theory, this result is stated as the noisy channel coding theorem [1], [2], [6], [7] .

In contrast to deflection routing, nonblocking routing in the Clos network will avoid internal contentions completely by conflict-free route assignments according to input requests [8]–[10]. The route assignment algorithms are based on the matching of bipartite graphs. In the error-correcting code, the low density parity check (LDPC) code developed by Gallager can also be represented by a bipartite graph [11], [12], and a subclass called the expander code was constructed by Sipser and Speilman [13]. The condition on the expander graph is a generalization of Hall's condition on complete matching, which suggests the resemblance between nonblocking route assignments and error-correcting codes. An extension of the Sipser and Speilman decoding algorithm of expander codes to route assignments of Benses networks is given to illustrate their correspondence.

| Packet Switched Clos Network | Transmission Channel |
|---|---|
| ∘ Random routing | ∘ Noisy channel capacity theorem |
| ∘ Deflection routing | ∘ Noisy channel coding theorem |
| ∘ Route assignment | ∘ Error-correcting code |
| ∘ BvN decomposition | ∘ Sampling theorem |
| ∘ Path switching | ∘ Noiseless channel |
| ∘ Scheduling | ∘ Noiseless coding theorem |

Fig. 3. Analogies between packet switching and information transmission

***The noiseless channel model of the Clos network***. A drastically different approach, called *path switching*, to deal with contention in a Clos network is proposed in [14]. A set of connection patterns of central modules is determined by the traffic matrix decomposition for this purpose. Path switching periodically uses this finite set of predetermined connections in the central stage of the Clos network to avoid online computation of route assignments. Once the connection patterns of central modules are fixed in each time slot, incoming packets can be scheduled accordingly in the input buffer. Regarding the predetermined connections as a code book, scheduling is a process similar to source coding in digital transmission [1], [2], [6], [7]. The smoothness of scheduling, like noiseless coding, is bounded by entropy inequalities.

The decomposition of traffic matrices, sometimes called the Birkhoff-von Neumann decomposition, has been widely used in SS/TDMA satellite communications [15], [16]. Path switching adopted this scheduling scheme in packet switching systems to guarantee the capacity of virtual paths in Clos networks. The same approach was applied to the crossbar switch, also called the Birkhoff-von Neumann switch, in [17]. Mathematically, the series expansion and reconstruction of doubly stochastic capacity matrices are similar to the Fourier series expansion and interpolation of sampling theorem of bandlimited signals [18]–[21]. They also serve the same function in terms of complexity reduction of communication systems. The capacity matrix decomposition employed in path switching will reduce the dimension of permutation space of a Clos network from $N!$ to $O(N^2)$ or even lower, while the sampling theorem in transmission reduces the infinite dimensional signal space of any duration $T$ to a finite number of samples.

The remainder of the paper is organized as follows. In section II, we propose the definition of pseudo signal-to-noise ratio (PSNR) of packet switch, and prove that the trade-off between the bandwidth and PSNR of a Clos network is the same as that given by noisy channel capacity theorem in transmission. In section III, we show that the loss probability of the Clos network with deflection routing is similar to the exponential error probability of binary symmetric channels with random coding. In section IV, we demonstrate the correspondence between nonblocking route assignments and error-correcting codes. In section V, we address the capacity allocation and capacity matrix decomposition issues related to the path switching implemented on a Clos network. In section VI, the entropy inequalities of smoothness of scheduling are derived, and comparisons of scheduling algorithms are discussed. Finally, the conclusion and future research are summarized in section VII.

## II. Parallel Characteristics of Contention and Noise

The apparent causes of contention and noise are quite different, but they both limit the performance of communication systems. In transmission, the trade-off between bandwidth and signal-to-noise ratio is given by Shannon-Hartley's noisy channel capacity theorem [19]. In packet switching, the difference between offered load and carried load reflects the degree of contention. In order to provide a common ground to compare contention and noise, a pseudo signal-to-noise ratio is defined to represent the carried load of a packet switch. We show that the packet-switched Clos network with random routing can be mathematically modelled as a noisy channel.

### A. Pseudo Signal-to-Noise Ratio of Packet switch

Output port contention occurs in a crossbar switch when several packets are destined for the same output in a time slot. Packets lost in contention will be dropped as shown in Fig. 4. Consider an $N \times N$ crossbar switch, without any prior knowledge of input traffic. We assume that input loading $\rho$ is homogeneous and output address is uniformly distributed, then the carried load $\rho'$ is the probability that an output is busy in any given time slot

$$\rho' = 1 - (1 - \frac{\rho}{N})^N \overset{N \to +\infty}{\longrightarrow} 1 - e^{-\rho} \tag{1}$$

The difference between offered load and carried load is caused by packet dropping due to contention at the outputs. Benes considered the number of calls in progress as the energy of the connecting network in the thermodynamics theory of traffic in a telephone system [22]. A similar traffic model was also explored in the broadband network [23]. If we regard packets as energy quantum, this proposition can be adopted in packet switching and stated as follows.

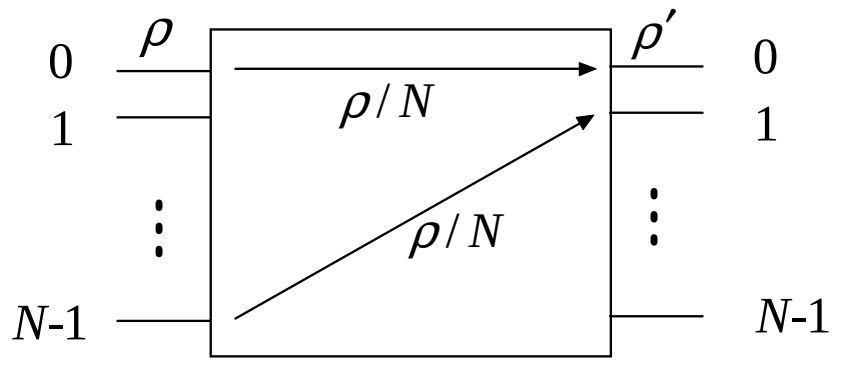

Fig. 4. The carried load of an $N \times N$ crossbar switch

***Proposition** 1:* (*Benes*): The *signal power* $S_p$ of an $N \times N$ crossbar switch is the number of packets carried by the system, and the *noise power* equals to $N_p = N - S_p$. Furthermore, both the signal power and noise power are Gaussian random variables when $N$ is large.

This proposition is verified by the Boltzmann model of packet distribution given in Appendix I [24]. It follows that the carried load of a crossbar switch can be converted to *pseudo signal-to-noise ratio*, a concept that is comparable to the signal-to-noise ratio (SNR) in transmission.

***Definition** 1:* The pseudo signal-to-noise ratio (PSNR) of a crossbar switch is the ratio of mean signal power to mean noise power

$$\text{PSNR} = \frac{E[S_p]}{E[N_p]} = \frac{N\rho'}{N - N\rho'} = \frac{\rho'}{1-\rho'} \tag{2}$$

It should be noted that the ratio of signal power to noise power SNR in transmission is defined by the ratio of the second moment of signal to that of noise, but PSNR is the ratio of their mean power here. Next, we will investigate the trade-off between bandwidth and PSNR of the Clos network.

### B. Clos Network with Random Routing as A Noisy Channel

In a three-stage Clos network, each pair of adjacent switch modules are interconnected by a unique link, and each module is a crossbar switch. As shown in Fig. 5, the $N \times N$ Clos network with $k$ input/output modules and $m$ central modules is denoted by $C(m, n, k)$ [5], [8], where $n$ is the number of input ports or output ports of each input module or output module, where $N = kn$. The dimensions of the modules in the input, middle and output stages are $n \times m$, $k \times k$ and $m \times n$, respectively. The Clos network $C(m, n, k)$ has the following characteristics:

1) Any central module can only assign to *one* input of each input module, and *one* output of each output module.
2) Source address $S$ and destination address $D$ can be connected through any central module.
3) The number of alternative paths between input $S$ and output $D$ is equal to the number of central modules $m$.

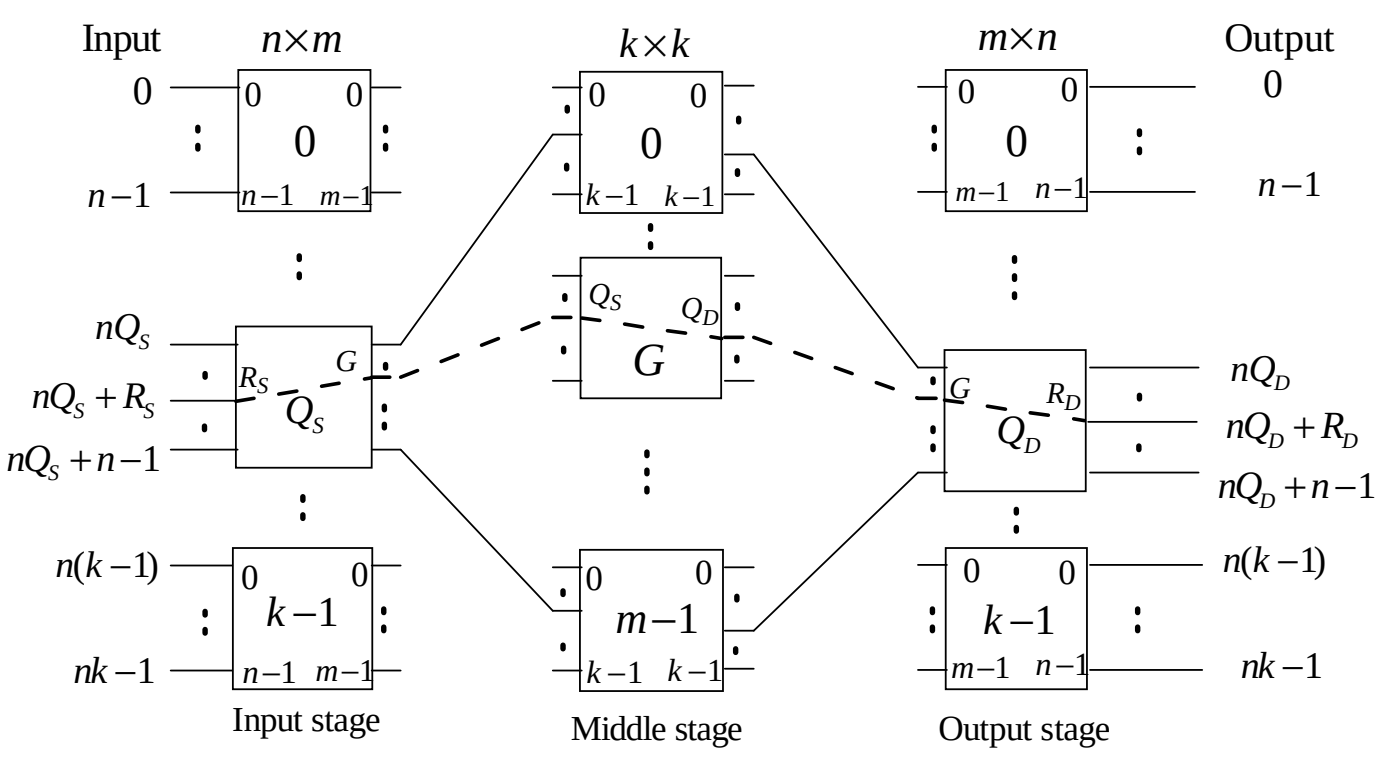

Fig. 5. The Clos network C$(m, n, k)$

In circuit switching theory, it is known that the Clos network $C(m, n, k)$ is rearrangeably nonblocking if $m \geq n$ and strictly nonblocking if $m \geq 2n - 1$ [8], [25]. Consider the number of cental modules $m$ as the bandwidth, the trade-off between bandwidth $m$ and PSNR of $C(m, n, k)$ with nonblocking routing is given by

$$\frac{E[S_p]}{E[N_p]} = \frac{n}{m - n} \tag{3}$$

In a packet switched Clos network, the maximal data rate of each input module is $n$ packets per time slot. Thus we define:

***Definition** 2:* The ratio $\sigma = n/m$ is the *maximal utilization* of a Clos network $C(m, n, k)$.

The route of each packet can be expressed by the numbering scheme of the Clos network [26]. The switch modules in each stage of the network, as well as the links associated with each module, are independently labelled from top to bottom. According to this numbering scheme, the source address $S$ is represented by the 2-tuple $S(Q_S, R_S)$, indicating the source $S$ is the link $R_S$ of the input module $Q_S$, where $R_S = [S]_n$ and $Q_S = \lfloor S/n \rfloor$ are the remainder and quotient of $S$ divided by $n$. Similarly, the destination address $D$ can be represented by the 2-tuple $D(Q_D, R_D)$. The path of an input packet from source $S$ to destination $D$ is determined by the choice of central module. Suppose the path goes through the central module $G$, then the routing tag is $(G, Q_D, R_D)$ and the path is expressed by:

$$S(Q_S, R_S) \rightarrow Q_S \stackrel{G}{\rightarrow} G \stackrel{Q_D}{\rightarrow} Q_D \stackrel{R_D}{\rightarrow} D(Q_D, R_D)$$

In a packet switched Clos network $C(m, n, k)$, contentions may occur in the middle stage even if destination addresses of input packets are all different. With random routing scheme, the trade-off between bandwidth $m$ and PSNR is given as follows:

***Theorem** 1:* The maximal data rate of each input module of the packet switched Clos network $C(m, n, k)$ with random routing is given by

$$n = m \ln(1 + \frac{E[S_p]}{E[N_p]}) \tag{4}$$

*Proof:*

The maximal data rate is achieved when the input loading $\rho = 1$ and destination addresses of input packets are all different in every time slot, in which case input modules and output modules are contention-free. If the central module is selected randomly for each input packet, then the loading on each input link of a central module is given by

$$\sigma = n/m$$

Since each switch module of the Clos network is a crossbar switch, the carried load on each output link of a central module is

$$\sigma' = 1 - (1 - \frac{n/m}{k})^k \stackrel{k \rightarrow +\infty}{\longrightarrow} = 1 - e^{-n/m} \tag{5}$$

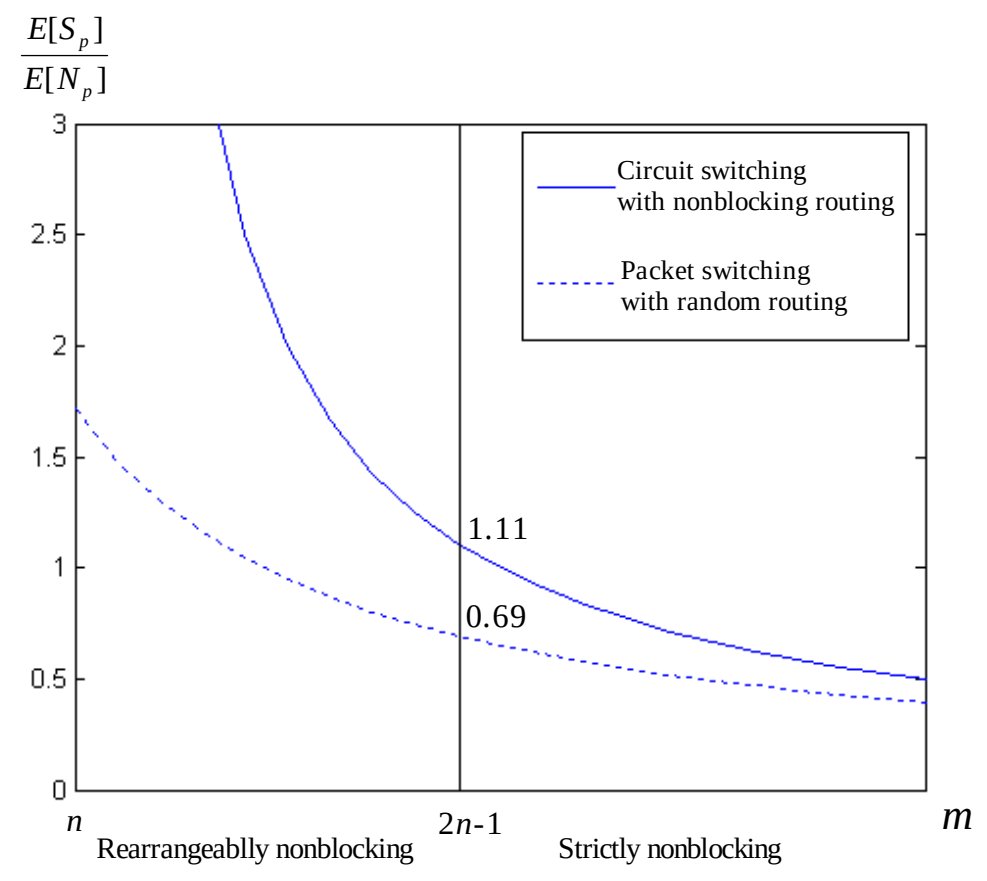

Fig. 6. Trade-off between bandwidth $m$ and PSNR of C$(m, n, k)$

Substituting the following PSNR for the carried load $\sigma'$ in (5)

$$\frac{E[S_p]}{E[N_p]} = \frac{km\sigma'}{km - km\sigma'} = \frac{\sigma'}{1 - \sigma'}$$

and taking logarithm, we obtain (4). □

The trade-offs given in (3) and (4) for constant data rate $n$ packets per time slot of each input/output module are plotted in Fig. 6, which shows the improvement of PSNR that can be achieved by contention-free routing. The same kind of trade-off between bandwidth and SNR in the transmission channel is stated in the Shannon-Hartley theorem [19] on noisy cannel capacity as follows:

***Noisy Channel Capacity Theorem*** *The channel capacity of a bandlimited Gaussian channel in the presence of additive Gaussian noise is given by*

$$C = W \log(1 + \frac{S}{N}) \tag{6}$$

*where $C$ is the capacity in bits per second, $B$ is the bandwidth of the channel in Hertz, and $\frac{S}{N}$ is the signal-to-noise ratio.*

With the understanding that the contention is mathematically analogous to noise, the rest of paper seeks to investigate the connections between the various routing schemes of the Clos network and the corresponding coding schemes of the transmission channel. We first demonstrate the similarity between deflection routing and random coding in the next section.

## III. Clos Network with Deflection Routing

Packet contention is inevitable in packet switching. One way of solving this problem without having to buffer the losing packet is to use deflection routing. In contrast to "store and forward" routing where the network allows buffering, deflection routing, also known as hot potato routing, is a routing scheme without buffering and can only be implemented for packet-switched networks [27]. In case the individual communication links cannot support more than one packet at a time, excessive packets will be transferred to other available links. Deflection routing actually utilizes idle or redundant links of the network as temporary storages. Redundancy is built into the switch design so that deflected packets can use extra stages to correct the deviated routes.

### A. Cascaded Clos Network

A cascaded Clos network is constructed by a sequence of alternate $C(n, n, k)$ and $C(k, k, n)$, as illustrated in Fig. 7. Each output link of a switch module is connected to a module in the next stage and an output concentrator (not shown in Fig. 7). A packet will be sent to the concentrator if its destination address matches the output numbering, otherwise it will continue with the remaining journey. The loss probability can be made arbitrarily small by providing a large enough number of stages.

In the cascaded Clos network, the destination address $D$ of a packet can either be expressed by $D = nQ_1 + R_1$ in the $C(n, n, k)$ network, or $D = kQ_2 + R_2$ in the $C(k, k, n)$ network, as shown in the example given in Fig. 7, and therefore the routing tag in the packet header comprises a 4-tuple $(Q_1, R_1, Q_2, R_2)$. A packet can reach its destination $D$ in any two consecutively successful steps, either $(Q_1, R_1)$ or $(Q_2, R_2)$. The Markov chain shown in Fig. 8 describes the complete journey of a packet, where $p_i$ and $q_i = 1 - p_i$ are respective probabilities of success and deflection in $C(n, n, k)$ and $C(k, k, n)$.

### B. Analysis of Deflection Clos Network

For the sake of simplicity, we assume that the cascaded Clos network is formed by $C(n, n, n)$ network uniformly. Since we will consider the worst case deflection probability when both $n$ and $k$ are sufficiently large, the assumption that $n = k$ does not lose the generality of analysis, and the packet loss probability will be a conservative estimate. Under this presumption, a simplified Markov chain is depicted in Fig. 9.

The probability of success $p$ in an $n \times n$ switch module is a function of input loading $\rho$. Under the homogeneous traffic assumption, we have

$$1 - p\rho = (1 - \frac{\rho}{n})^n \xrightarrow{n\to\infty} e^{-\rho}.$$

Thus, the worst probability of success is given by

$$p = \frac{1 - e^{-\rho}}{\rho}$$

The probabilities of success $p$ and deflection $q = 1 - p$ versus the loading $\rho$ are plotted in Fig. 10(a). Let $G_i(k)$ be the probability that the packet in state $i$ will reach the output state $O$ in exactly $k$ steps. The following equations can be derived from the Markov chain shown in Fig.9.

$$G_O(k) = \begin{cases} 1 & \text{if} \quad k = 0 \\ 0 & \text{otherwise.} \end{cases}$$

$$G_R(k) = pG_0(k-1) + qG_Q(k-1), \qquad k = 1, 2, \ldots$$

$$G_Q(k) = pG_R(k-1) + qG_Q(k-1), \qquad k = 1, 2, \ldots \quad (7)$$

The generating function $G_Q(k)$ is given by

$$G_Q(z) = \sum_{k=0}^{\infty} G_Q(k) z^k = \frac{p^2 z^2}{1 - qz - pqz^2}, \quad (8)$$

from which we obtain the following steady state probabilities

$$G_Q(k) = \begin{cases} 0 & k = 0, 1 \\ \frac{p}{vq}(pq)^{(k+1)/2} \cosh(k-1)\theta & k = 2, 4, 6 \cdots \cdot \\ \frac{p}{vq}(pq)^{(k+1)/2} \sinh(k-1)\theta & k = 3, 5, 7 \cdots \cdot \end{cases} \quad (9)$$

where $v = \frac{\sqrt{q^2+4pq}}{2}$ and $\theta = \ln \frac{q+\sqrt{q^2+4qp}}{2\sqrt{pq}}$. For given network length $L$, a conservative estimate of packet loss probability is given as follows.

$$P_{loss} \leq \sum_{k=L+1}^{\infty} G_Q(k), \quad (10)$$

which can be explicitly expressed

$$P_{loss} \leq \begin{cases} \frac{1}{vq}(pq)^{(L+2)/2} \sinh(L+2)\theta & \text{for even length } L\ ; \\ \frac{1}{vq}(pq)^{(L+2)/2} \cosh(L+2)\theta & \text{for odd length } L. \end{cases}$$

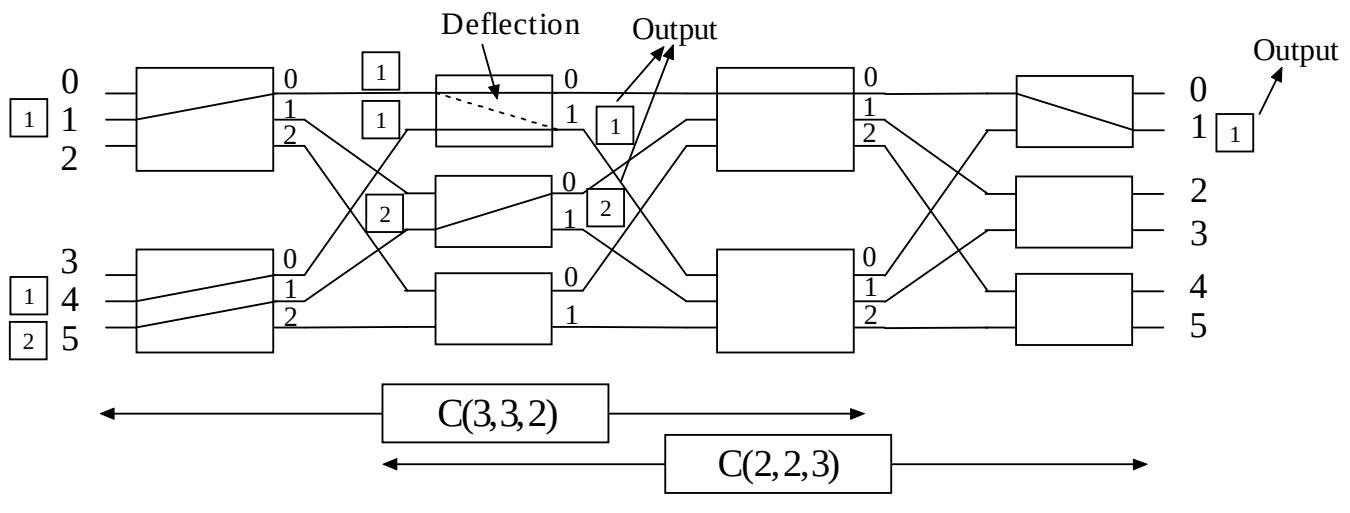

| Routing coding in C(3,3,2) | | |
|---|---|---|
| Destination | $Q_1$ | $R_1$ |
| 0 | 0 | 0 |
| 1 | 0 | 1 |
| 2 | 0 | 2 |
| 3 | 1 | 0 |
| 4 | 1 | 1 |
| 5 | 1 | 2 |

D = $3Q_1$+$R_1$

| Routing coding in C(2,2,3) | | |
|---|---|---|
| Destination | $Q_2$ | $R_2$ |
| 0 | 0 | 0 |
| 1 | 0 | 1 |
| 2 | 1 | 0 |
| 3 | 1 | 1 |
| 4 | 2 | 0 |
| 5 | 2 | 1 |

D = $2Q_2$+$R_2$

Fig. 7. Cascaded Clos network with deflection routing

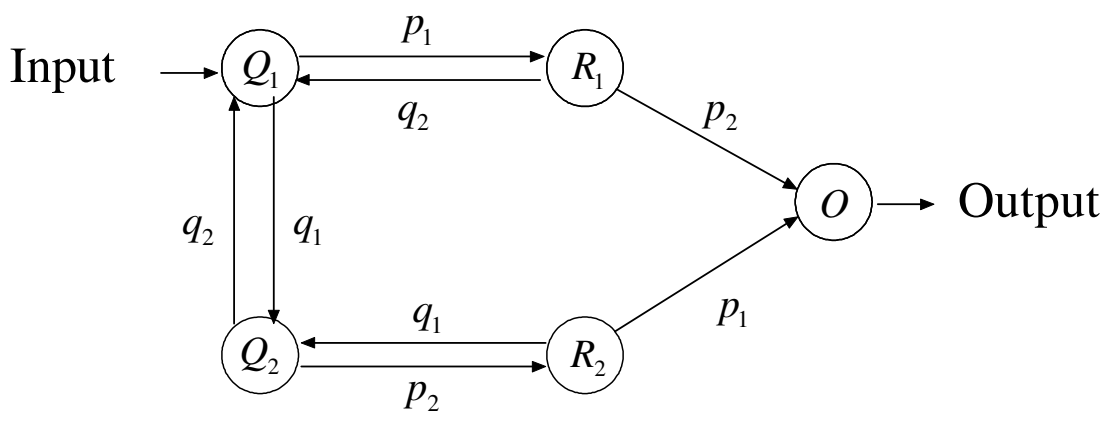

Fig. 8. The Markov chain of deflection routing

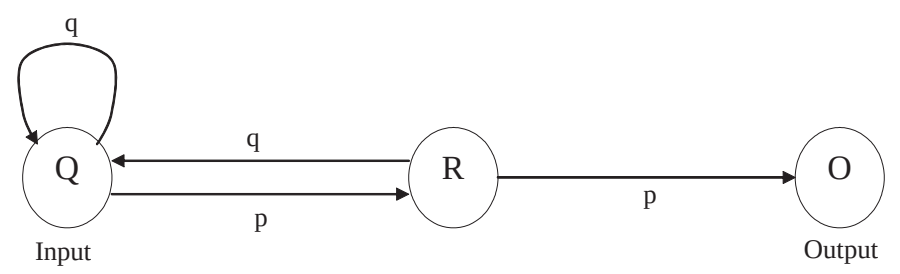

Fig. 9. The simplified Markov chain of deflection routing

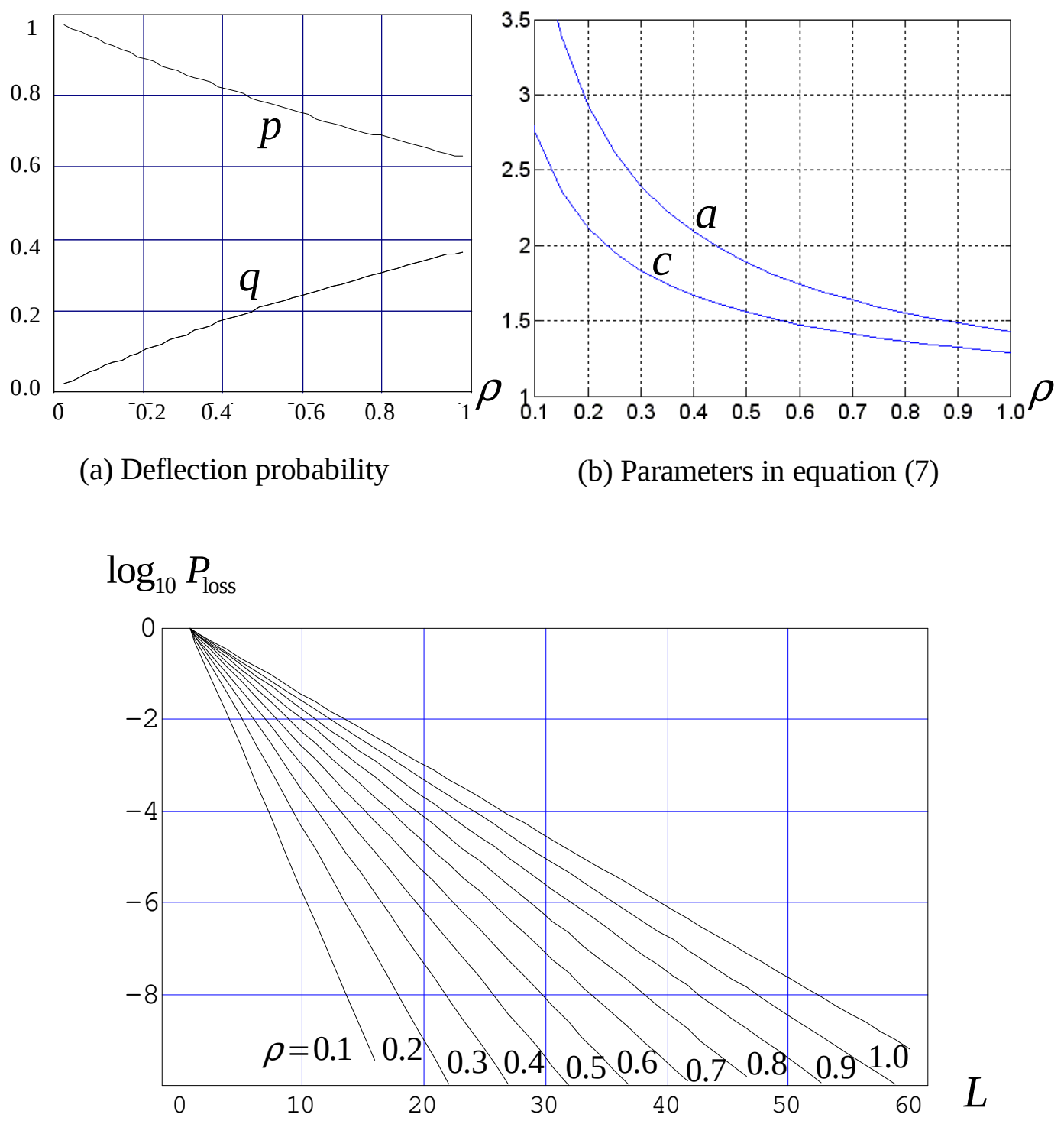

(a) Deflection probability (b) Parameters in equation (7)

(c) Loss probabilities versus network length

Fig. 10. Loss probabilities of deflection routing

When $L$ is large, the logarithm of $P_{loss}$ is linear in $L$ as shown by the curves in Fig.10(c)

$$\ln P_{loss} \le m(L+2) + b, \tag{11}$$

where $m = \ln(\frac{q+\sqrt{q^2+4pq}}{2})$ and $b = \ln(\frac{1}{q\sqrt{q^2+4pq}})$. In the worst case when $\rho = 1$, $p = 0.6321$, $q = 0.3679$, $m = -0.3566$, and $b = 0.9683$.

***Theorem** 2:* If the offered load $\rho \le 1$ on each input of the Clos network with deflection routing, then the loss probability can be arbitrarily small and the carried load $\rho'$ on each output can be arbitrarily close to the carried load $\rho$.

*Proof:* It follows from (11) that the loss probability in terms of offered load $\rho$ and carried load $\rho'$ is bounded by

$$P_{loss} = \frac{\rho - \rho'}{\rho} \le ca^{-L}. \tag{12}$$

The two parameters $a$ and $c$ are given by following functions of deflection probability $q$

$$a = \frac{2}{q + \sqrt{q^2 + 4pq}} > 1,$$

and

$$c = \frac{(q + \sqrt{q^2+4pq})^2}{4q\sqrt{q^2+4pq}} > 1.$$

The parameters $a$ and $c$ versus the offered load $\rho$ are plotted in Fig. 10(b), in which $a = 1.4285$ and $c = 1.2906$ when $\rho = 1$. □

Conversely, if the offered load $\rho > 1$, then the loss probability is unbounded, because the number of packets input to a switch module could be more than the total number of outputs and there is not enough space to deflect packets, which can be lost halfway through.

Compare the deflection Clos network with the binary symmetric channel (BSC) with random coding, as illustrated in Fig. 11. Table I provides the parallels between the above theorem and Shannon's main theorem [1], [2], [6], [7] stated as follows.

***Noisy Channel Coding Theorem*** *Given a noisy channel with capacity $C$ and information transmitted at a rate $R$, then if $R \le C$, there exists a coding technique which allows the probability of error at the receiver to be made arbitrarily small. This means it is possible to transmit information without error up to a limit $C$. However, if $R > C$, the probability of error at the receiver increases without bound.*

The capacity of BSC with cross probability $q = 1 - p$ is given by

$$C = 1 + p\log p + (1-p)\log(1-p) \tag{13}$$

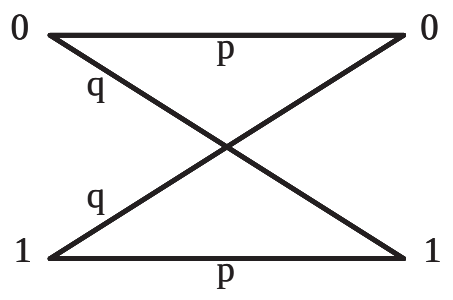

Fig. 11. Binary Symmetric Channel (BSC)

The noisy channel coding theorem states that there exists an encoding function $E : \{0,1\}^k \to \{0,1\}^n$ and a decoding function $D : \{0,1\}^n \to \{0,1\}^k$, such that the error probability that the receiver gets a wrong message is bounded by

$$P_e \le a^{-n} + c^{-n} \qquad a > 1, c > 1 \tag{14}$$

if the code rate $R = \frac{k}{n} = C - \delta$ for some $\delta > 0$ [28]. Thus, the error probability $P_e$ can be arbitrarily small for a sufficiently large $n$. Conversely, if the rate $R = \frac{k}{n} = C + \delta$ for some $\delta > 0$, then the error probability is unbounded.

| Deflection Clos Network | Binary Symmetric Channel |
|---|---|
| Deflection Probability $q < \frac{1}{2}$ | Cross Probability $q < \frac{1}{2}$ |
| Deflection Routing | Random Coding |
| $\rho \le 1$ | $R \le C$ |
| Exponential Loss Probability $P_{loss} \le ca^{-L}$ | Exponential Error Probability $P_e \le a^{-n} + c^{-n}$ |
| Complexity increases with network length $L$ | Complexity increases with code length $n$ |
| Equivalent set of outputs | Typical set decoding |

TABLE I

COMPARISON OF DEFLECTION CLOS NETWORK AND BINARY SYMMETRIC CHANNEL

The similar behavior of the deflection routing of the Clos network and the random coding of noisy channel suggests the connection between nonblocking route assignments and error-correcting codes. This analogy can be extended to multistage interconnection networks as well. The deflection routing of the tandem banyan network proposed in [29] is equivalent to the concept of error-detection and retransmission, while the deflection routing of the dual-shuffle exchange network is a distributed error-correction algorithm [30]. These points will be further elaborated on in the next section.

## IV. Route Assignments and Error-Correcting Codes

The Clos network $C(m,n,k)$ is rearrangeable if $m \geq n$, meaning that it can realize connections of any permutations between inputs and outputs with the possibility of rearranging the calls in progress. Both Slepian-Duguid's nonblocking theorem [8], [25] and M. C. Paull's rearrangeable theorem [31] on route assignments of the Clos network are rooted in matching theory, which starts with Hall's marriage theorem that gives the existence condition of a complete matching in a given bipartite graph [32]. In the error-correcting code, the low density parity check (LDPC) code can also be represented by a bipartite graph [11], [12]. The connection between route assignment and graphic code in the context of bipartite graph is the main point addressed in this section.

### *A. Complete Matching in Bipartite Graphs*

A *bipartite graph* $G = (V_L, V_R, E)$ consists of two finite sets $V_L$, $V_R$ of vertices, and a collection of edges $E$ connecting vertices in $V_L$ to vertices in $V_R$.

***Definition** 3:* A *complete matching* in a bipartite graph $G$ is an injective function $f: V_L \to V_R$ so that for every $x \in V_L$, there is an edge in $E$ whose endpoints are $x$ and $f(x)$.

***Definition** 4:* The *neighborhood* of any subset $A \subset V_L$ is defined by

$$N_A = \{b|(a,b) \in E, a \in A\} \subseteq V_R$$

which are endpoints of an edge in $E$ whose other endpoints lie in $A$.

The necessary and sufficient condition for a bipartite graph to have a complete matching is given by Hall's marriage theorem [32], [33]. Because it is the most fundamental theorem in switching theory, we state the theorem and give R. Rado's elegant proof [34] below.

***Theorem** 3: (Hall)* Let $G = (V_L, V_R, E)$ be a bipartite graph, there exists a complete matching for $G$ if and only if

$$|N_A| \geq |A| \tag{15}$$

for any subset $A \subseteq V_L$.

*Proof:* It is clear that the condition is essential for the existence of a complete matching, it remains to show that it is sufficient. Let $V_L = \{1, 2, ..., n\}$, and $F = \{N_1, N_2, ..., N_n\}$, by way of contradiction, suppose $N_1$ contains $x, y$, the removal of either $x$ or $y$ violates Hall's condition. There exists $A, B \subseteq \{2, ..., n\}$ with the property

$$R_A = N_A \cup (N_1 - \{x\}), \qquad |R_A| \leq |A|$$

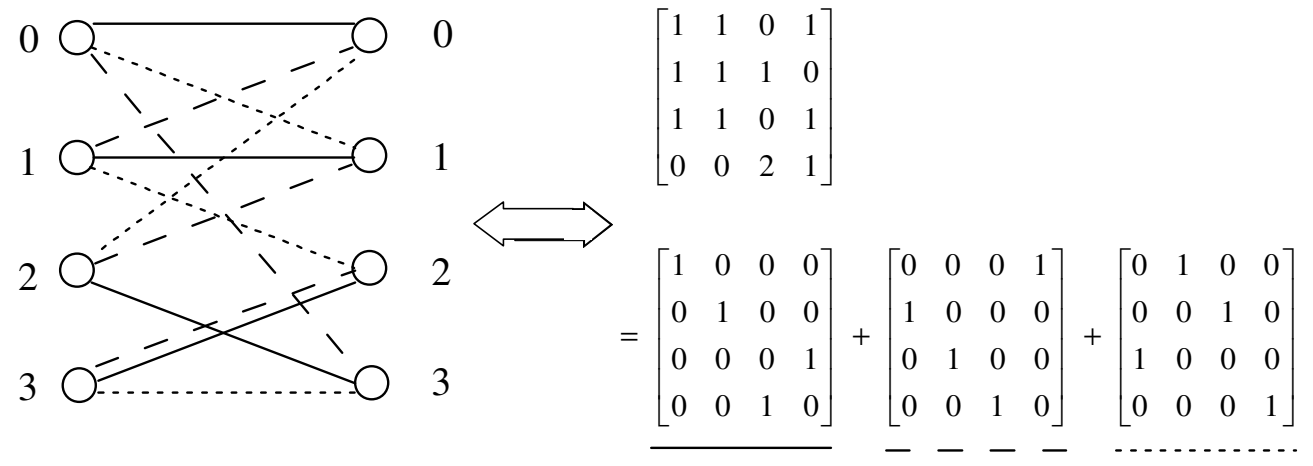

Fig. 12. An edge colored regular bipartite graph and decomposition of integer doubly stochastic matrix

$$R_B = N_B \cup (N_1 - \{y\}), \qquad |R_B| \leq |B|$$

Then $|R_A \cup R_B| = |N_{A\cup B} \cup N_1|$, and $|R_A \cap R_B| \geq |N_{A\cap B}|$. It follows that

$$\begin{aligned} |A| + |B| \geq |R_A| + |R_B| &= |R_A \cup R_B| + |R_A \cap R_B| \\ &\geq |N_{A\cup B} \cup N_1| + |N_{A\cap B}|. \end{aligned} \tag{16}$$

On the other hand, from Hall's condition, we have

$$|N_{A\cup B} \cup N_1| + |N_{A\cap B}| \geq |A \cup B| + 1 + |A \cap B| = |A| + |B| + 1. \tag{17}$$

Combining (16) and (17) together will lead to

$$|A| + |B| \geq |A| + |B| + 1,$$

a contradiction. Hence, the removal of either $x$ or $y$ does not violate Hall's condition, and a complete matching can be obtained by repeating this procedure until each $N_i$ contains only one element. □

The nonblocking route assignment of the Clos network and the edge coloring of the bipartite graph stated in the following theorem are equivalent consequences of Hall's marriage theorem.

***Theorem** 4:* The following statements are equivalent:

1. A regular bipartite graph $G$ with degree $n$ can be edge colored by $m$ colors if $m \geq n$.

2. *(Slepian-Duguid)* The Clos network $C(m,n,k)$ is rearrangeably nonblocking if $m \geq n$.

*Proof:* 1. For any subset $A \subseteq V_L$, the edges terminate on vertices in A must be terminated on vertices in its neighborhood set $N_A$ at the other end, Hall's condition holds because

$$n|N_A| \geq n|A| \Rightarrow |N_A| \geq |A|.$$

The bipartite graph $G$ can be reduced to $n$ complete matching by repeatedly applying Hall's theorem. It is clear that edge coloring can be satisfied if $m \geq n$. The edge colored bipartite graph is equivalent to the decomposition of an integer doubly stochastic matrix into permutation matrices. An example to illustrate this point is shown in Fig.12.

2. Consider a set of call requests $\{(S_0, D_0), \cdots, (S_{N-1}, D_{N-1})\}$, in which the destination addresses are all different, and suppose the central module $G_i$ is assigned to the request $(S_i, D_i)$ for $i = 0, \cdots, N-1$.

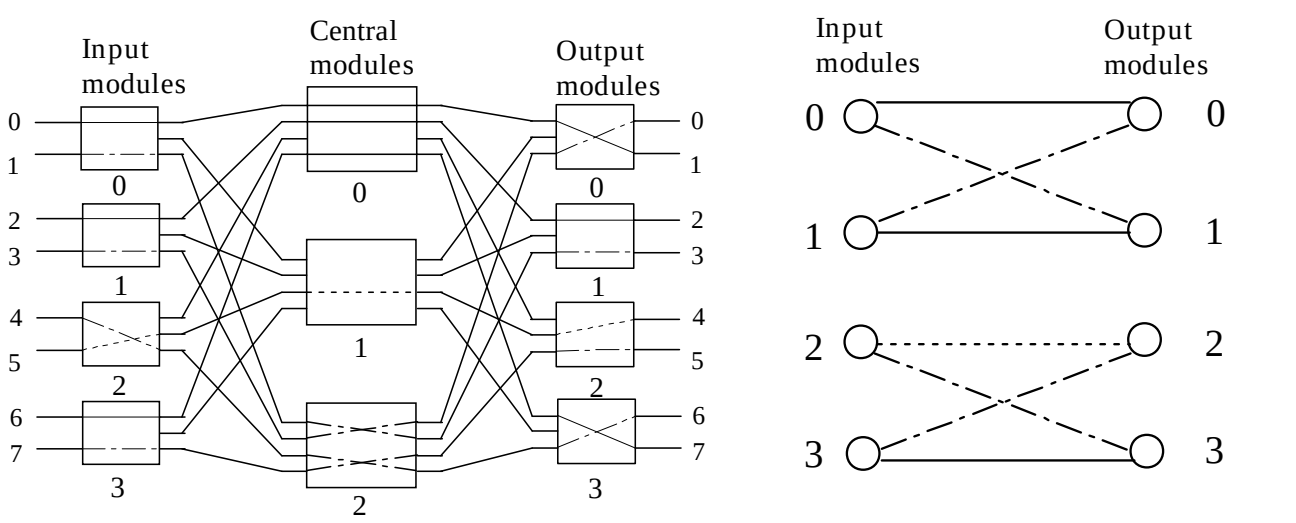

(a) Three-stage Clos network (b) Equivalent regular bipartite graph

Fig. 13. Correspondence between Clos network and bipartite graph

| S | 0 | 1 | 2 | 3 | 4 | 5 | 6 | 7 |
|---|---|---|---|---|---|---|---|---|
| D | 1 | 3 | 2 | 0 | 6 | 4 | 7 | 5 |
| G | 0 | 2 | 0 | 2 | 2 | 1 | 0 | 2 |
| Q | 0 | 1 | 1 | 0 | 3 | 2 | 3 | 2 |
| R | 1 | 1 | 0 | 0 | 0 | 0 | 1 | 1 |

TABLE II
THE SET OF ROUTING TAGS $(G, Q, R)$

This set of assignments is nonblocking, or contention-free, if and only if

$$\lfloor S_i/n \rfloor = \lfloor S_j/n \rfloor \Rightarrow G_i \neq G_j$$

and

$$\lfloor D_i/n \rfloor = \lfloor D_j/n \rfloor \Rightarrow G_i \neq G_j$$

for all $i \neq j$. It simply means that if either of the two sources $S_i, S_j$ are on the same input modules, or the two destinations $D_i, D_j$ are on the same output module, then the central modules assigned to $(S_i, D_i), (S_j, D_j)$ must be different.

The route assignments can be formulated as the edge coloring of a bipartite graph $G(V_L, V_R, E)$, in which vertices of $V_L$ represent input modules and vertices of $V_R$ represent output modules, and each connection request $(S_i, D_i)$ is represented by an edge $(\lfloor S_i/n \rfloor, \lfloor D_i/n \rfloor)$ in $E$. This bipartite graph is regular with degree $n$, and can be colored by $m$ colors if $m \geq n$, such that each color represents a central module and the edge coloring is the set of route assignments. □

An example to show the equivalence between route assignment of the Clos network and edge coloring of the bipartite graph is given in Fig.13. The resulting route assignments of the call request are listed in Table II. The connection between route assignment and error-correcting code will be addressed in the sequel.

### *B. Graphical Codes*

The low-density parity-check (LDPC) code is a class of linear block code that can be represented by bipartite graphs $G(V_L, V_R, E)$, called Tanner graph [12], in which $V_L$ is the set of variable-vertices and $V_R$ the set of constraint-vertices. The parity matrix specifies the edge set $E$. Each variable can assume the value of 0 or 1, and a constraint is satisfied if the sum of all the variables adjacent to it is zero mod 2. Fig. 14

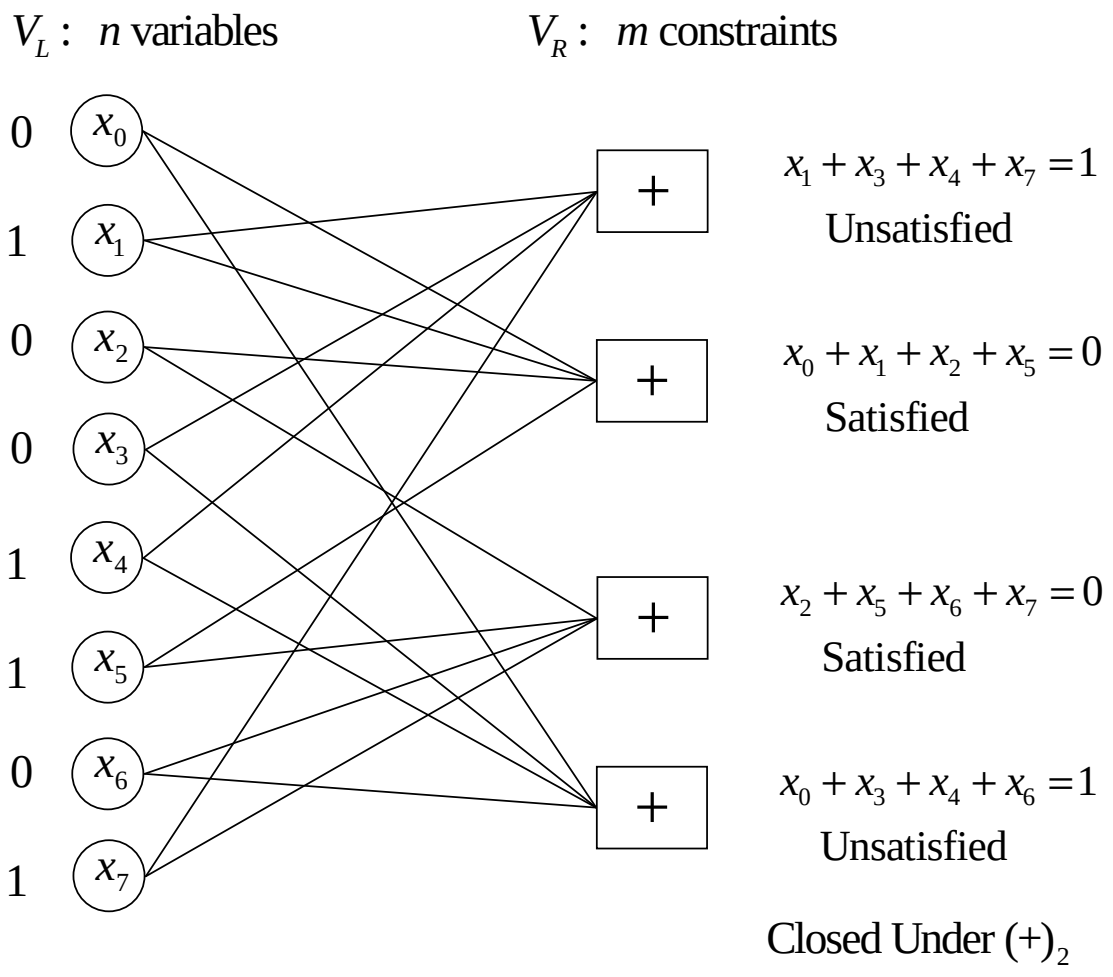

Fig. 14. Low density parity check code

illustrates an example of such a bipartite graph that represents the following matrix

$$\begin{bmatrix} 0 & 1 & 0 & 1 & 1 & 0 & 0 & 1 \\ 1 & 1 & 1 & 0 & 0 & 1 & 0 & 0 \\ 0 & 0 & 1 & 0 & 0 & 1 & 1 & 1 \\ 1 & 0 & 0 & 1 & 1 & 0 & 1 & 0 \end{bmatrix} \tag{18}$$

A vector $x \in \{0, 1\}^n$ is a code word if and only if $x$ satisfies all constraints. The set of all code words is closed with respect to mod 2 sum, and forms a linear subspace of $\{0, 1\}^n$.

A subclass of graphical codes based on expander graphs was constructed by Sipser and Speilman [13]. The expander code has the following linear time sequential decoding algorithm.

---

Sipser-Speilman decoding algorithm:
If there is a variable vertex $v$ such that most of its neighboring constraints are unsatisfied, flip the value of $v$. Repeat.

---

The decoding algorithm of expander codes guarantees that the number of unsatisfied vertices will be monotonically decreasing until all constraints are satisfied. The main result is stated in the following theorem.

***Theorem** 5: (Sipser-Speilman)* Let $G(V_L, V_R, E)$ be the bipartite graph of size $|V_L| = n, |V_R| = m$ that is $k-$regular on the left. Assume that for any subset $A \subseteq V_L$ of size $|A| \leq \alpha n$, $|N_A| > \frac{3k}{4}|A|$, then the decoding algorithm will correct up to $\frac{\alpha n}{2}$ errors.

The similarity between Hall's condition (15) on complete matching and the condition on expander graphs in the above theorem is quite obvious. The resemblance between these two structure conditions on bipartite graphs suggests the connection between error-correcting code and route assignment of Clos network. The Sipser-Speilman decoding algorithm is extended to the nonblocking route assignment of Benes network to illustrate their correspondence.

### *C. Route Assignments of Bense Network*

A Bense network is a multi-stage connecting network that can be recursively constructed from Clos network $C(2, 2, k)$

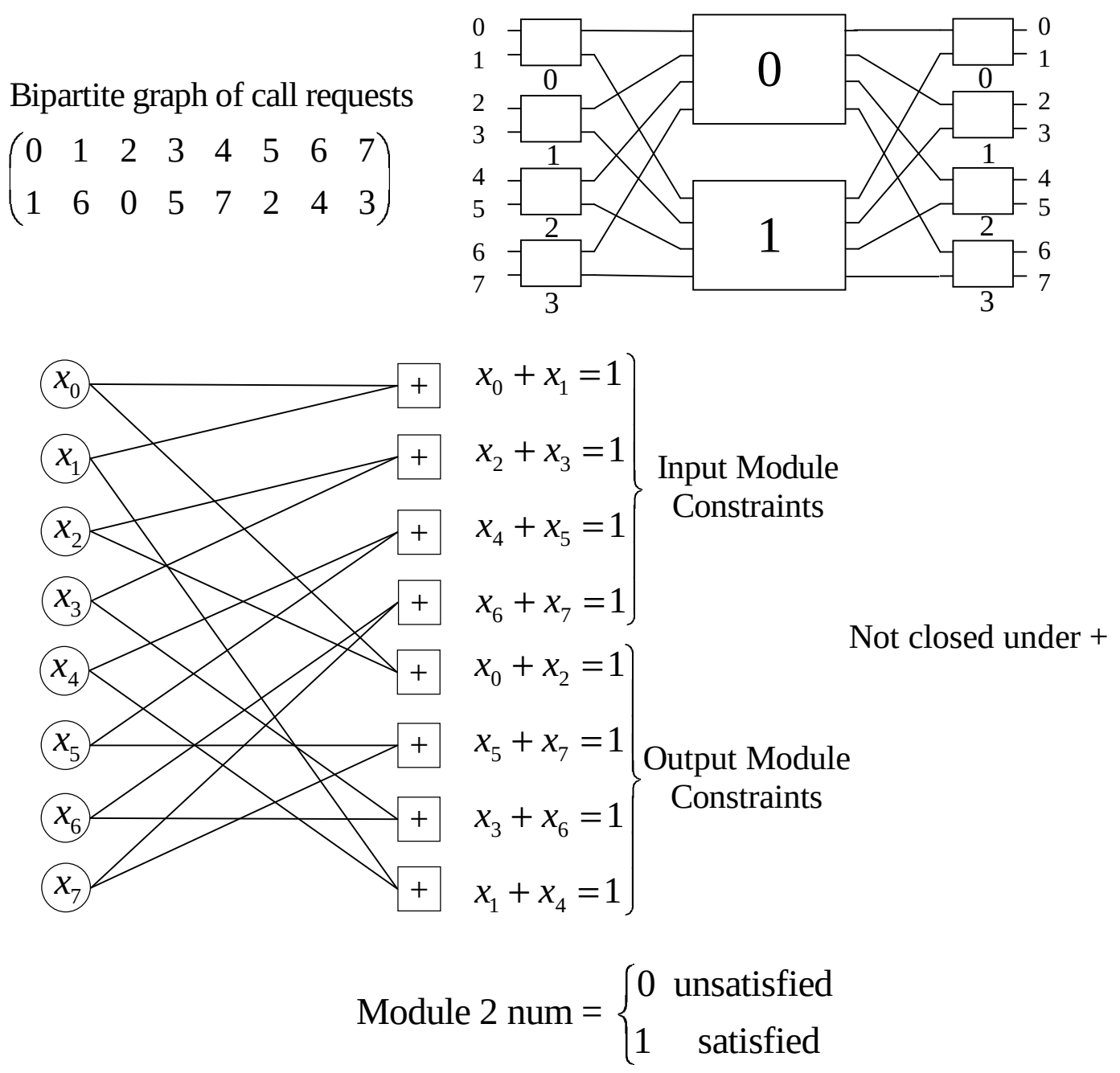

Fig. 15. Nonblocking constraints of Benes network

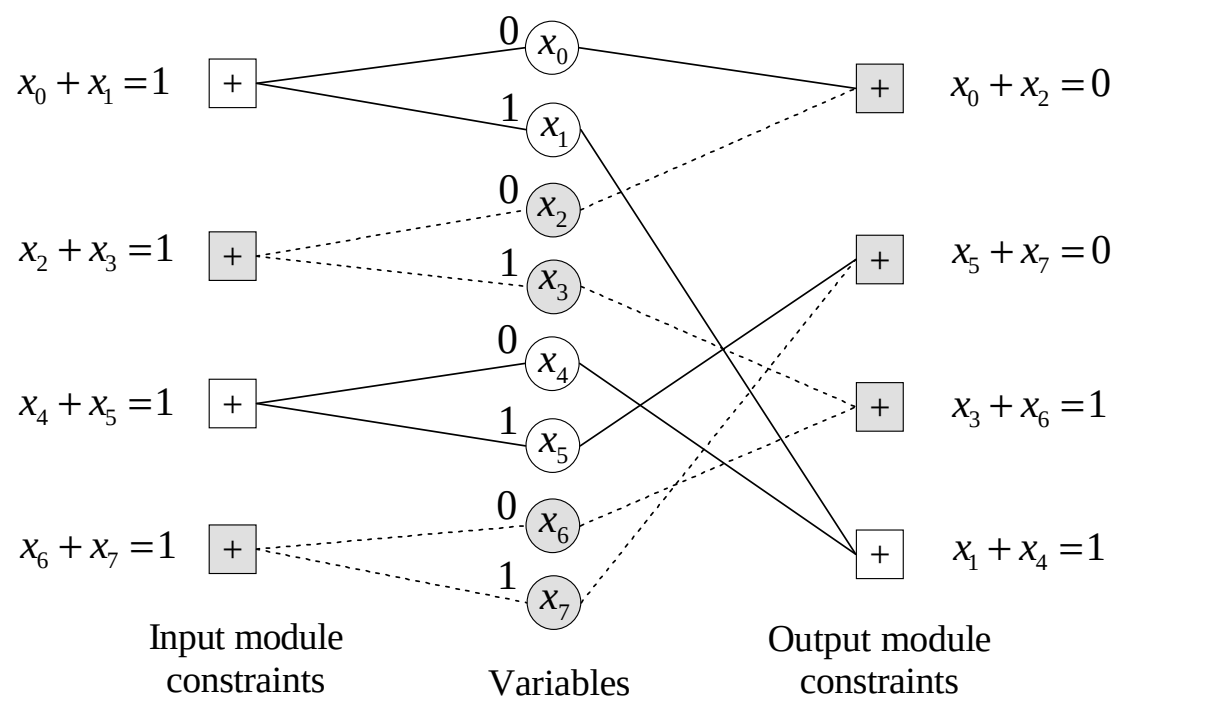

Fig. 16. The flip algorithm for nonblocking route assignments

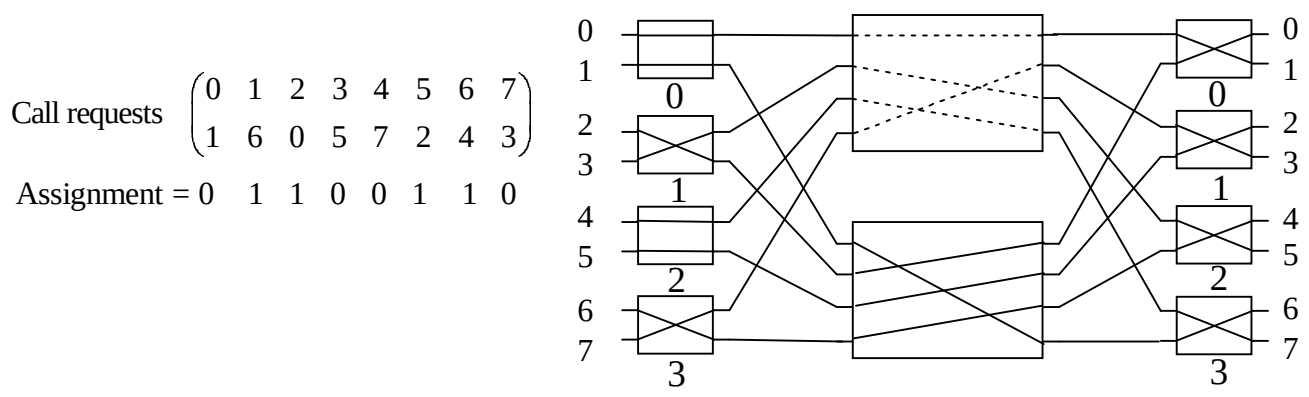

Fig. 17. The connection requests and and route assignments

[8], [25]. A set of call requests is a mapping from inputs to outputs. An example is given in (19), which displays a set of connection requests in an $8 \times 8$ Benes network. The upper row of the matrix is the ordered inputs from 0 to 7, and the lower row consists of the respective outputs.

$$\pi = \begin{pmatrix} 0 & 1 & 2 & 3 & 4 & 5 & 6 & 7 \\ 1 & 6 & 0 & 5 & 7 & 2 & 4 & 3 \end{pmatrix} \tag{19}$$

The two central modules in $C(2, 2, \frac{N}{2})$, as shown in Fig. 18(a), are labelled respectively by 0 and 1. Let $x_i$ be the binary variable associated with the request $(i, \pi(i))$, for $i = 0, \cdots, N-1$. The value of $x_i \in \{0, 1\}$ defines the following route assignment

$$x_i = \begin{cases} 0 & \text{if module 0 assigned to } (i, \pi(i)) \\ 1 & \text{otherwise} \end{cases} \tag{20}$$

Since the same central module cannot be assigned to two inputs or two outputs on the same module, the necessary and sufficient condition on nonblocking route assignment can be simply formulated as a set of linear constraints

$$x_i + x_j = 1 \tag{21}$$

for all $i \neq j$ such that either two inputs $i$ and $j$ are on the same input module, or two outputs $\pi(i))$ and $\pi(i))$ are on the same output module.

As shown in Fig.15, let $V_L = \{x_0, \cdots, x_{N-1}\}$ be the set of variables and $V_R$ be the set of constraints (21), a bipartite graph $G(V_L, V_R, E)$ similar to that of LDPC can be constructed to solve for the values of $x_i's$, which determine the nonblocking routes of the set of call requests in $\pi$. It should be noted that the set of solutions to (21) is not closed with respect to mod 2 sum. In fact, two solutions in each connected component of the bipartite graph $G(V_L, V_R, E)$ are complementary to each other, and the total number of solutions of (21) should be $2^g$, where $g$ is the number of connected components of $G$. The following route assignment algorithm is a modification of Sipser-Speilman decoding algorithm.

---

The flip algorithm for nonblocking route assignments:

- Step 1 Initially assign $x_0 = 0, x_1 = 1, x_2 = 0, x_3 = 1, \cdots$ to satisfy all input module constraints.
- Step 2 In each cycle of the bipartite graph, unsatisfied vertices divide the cycle into line segments. Label them $\alpha$ and $\beta$ alternately and flip the values of all variables located in $\alpha$ segments.

---

The initial values of all variables can be arbitrarily assigned. Compared with the Sipser-Speilman decoding algorithm, the initial assignment can be considered as the received signal, and errors are systematically removed by the flip algorithm to satisfy all constraints.

***Theorem** 6:* All constraints are satisfied when the flip algorithm terminates.

*Proof:* The proof of this theorem is illustrated by the example displayed in Fig. 16. First, we want to show that every cycle has an even number of unsatisfied vertices. It is clear that the total number of constraints is even because there is an equal number of input modules and output modules in each cycle. Assume by way of contradiction that the number of unsatisfied vertices is odd, then the number of satisfied vertices must also be odd. Let $u_0$ and $u_1$ be the number of unsatisfied vertices whose neighboring variables both equal 0 and 1, respectively, and let $s$ be the number of satisfied vertices. Then, the number of variables equal to 0 and 1, respectively, should be $u_0+\frac{s}{2}$ and $u_1+\frac{s}{2}$, which are impossible if $s$ is odd.

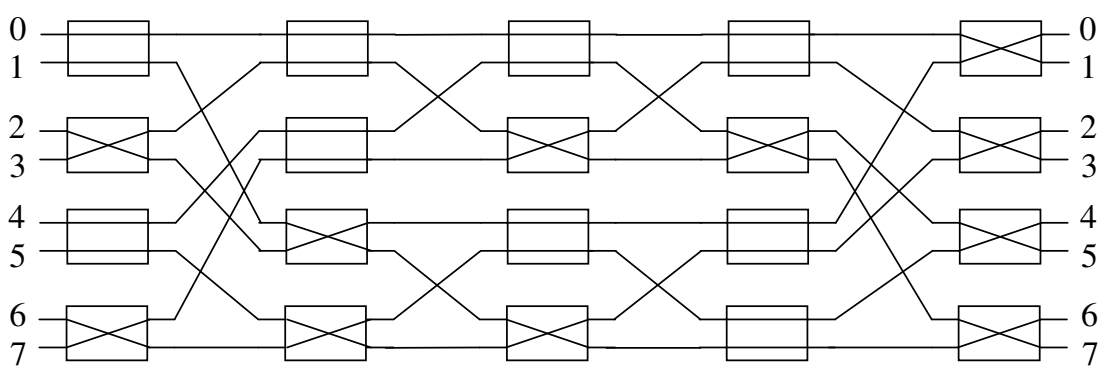

Fig. 18. The complete route assignments of Benes network

Next, we want to show that all constraints are satisfied when the flip algorithm terminates as shown in Fig.16. For an arbitrary constraint vertex $v \in V_R$, let $x_i, x_j \in V_L$ be the two neighboring variables of $v$. If $v$ is unsatisfied, the two variables $x_i$ and $x_j$ should have the same value but are located in segments bearing different labels. Only one located in the $\alpha$ segment will flip the value, and the vertex $v$ should be satisfied when the algorithm terminates.

On the other hand, if the constraint $v$ is initially satisfied, then $x_i$ and $x_j$ flip simultaneously if they are both in an $\alpha$ segment, or else they keep the same value if they are located in a $\beta$ segment. Hence, the constraint $v$ remains satisfied in either case when the algorithm terminates. □

The route assignments of the set of call requests (19) resulting from the flip algorithm are displayed in Fig.17, and the complete assignments shown in Fig. 18 can be determined iteratively. If both $N$ and $n$ are of the power of 2, the set of equations (21) can also be solved by the parallel algorithm proposed in [35] with time complexity on the order of $O(\log^2(N))$.

## V. Clos Network as Noiseless Channel-Path Switching

Path switching proposed in [14] is a compromise of static routing and dynamic routing schemes for a Clos network. A set of connection patterns of the central modules is predetermined according to the traffic between pairs of input and output modules. This set of connection patterns is repeatedly used in a cyclic manner such that online path hunting can be avoided and bandwidth requirements can be satisfied in the long run.

The scheduling of path switching is based on the following relationships between the edge colored bipartite graph and the connection pattern in the middle stage of the Clos network as depicted in Fig. 13 earlier:

1) The number of edges $e_{ij}$ represents the number of packets that can be sent from input module $I_i$ to output module $O_j$, the *virtual path* $V_{ij}$, in one time slot,
2) Each color of the bipartite graph corresponds to a central module of the Clos network.

This correspondence suggests that the scheduling of switch according to predetermined connection patterns if traffic matrix is expressed as weighted sum of a finite number of permutation matrices as shown in Fig. 19.

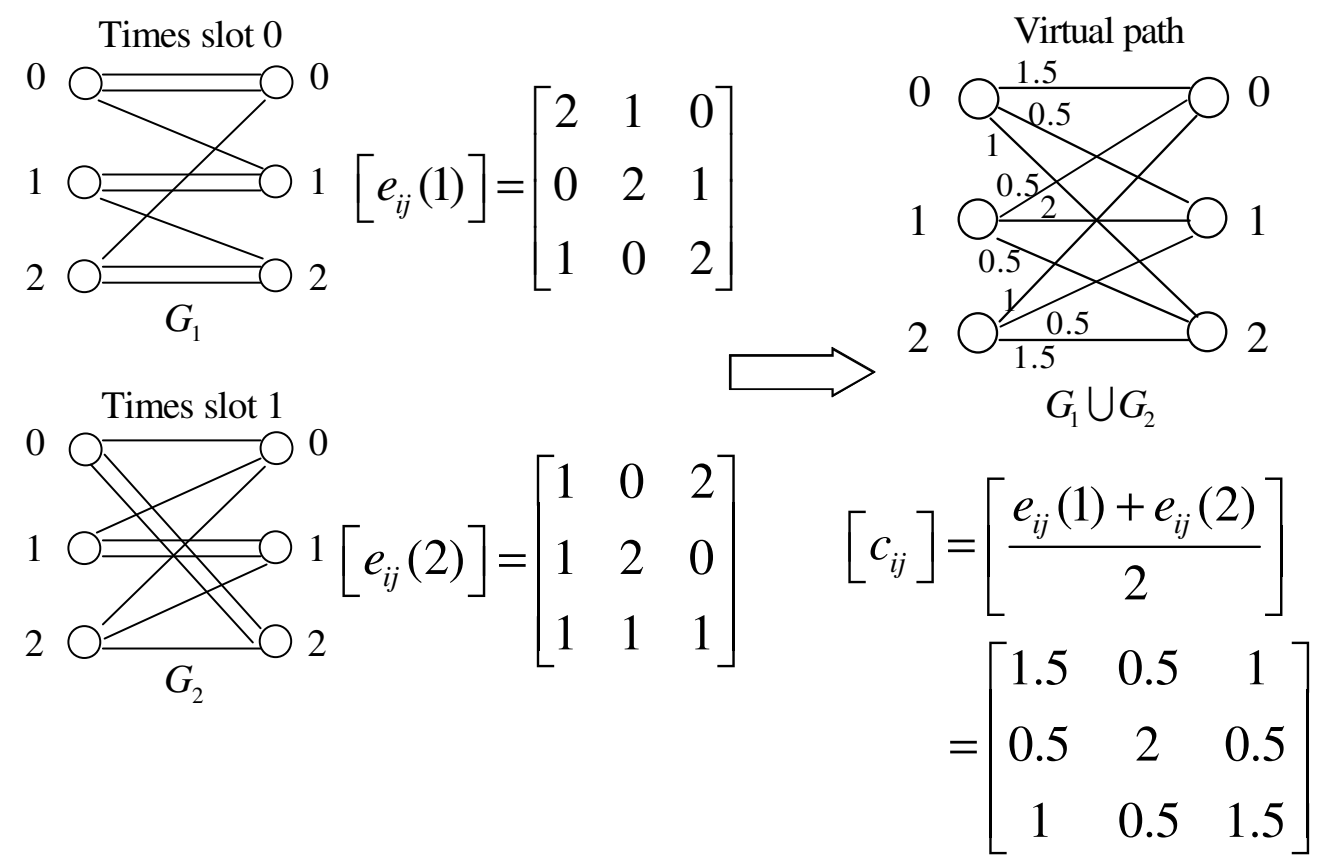

Fig. 19. Capacity of virtual path as average number of edges

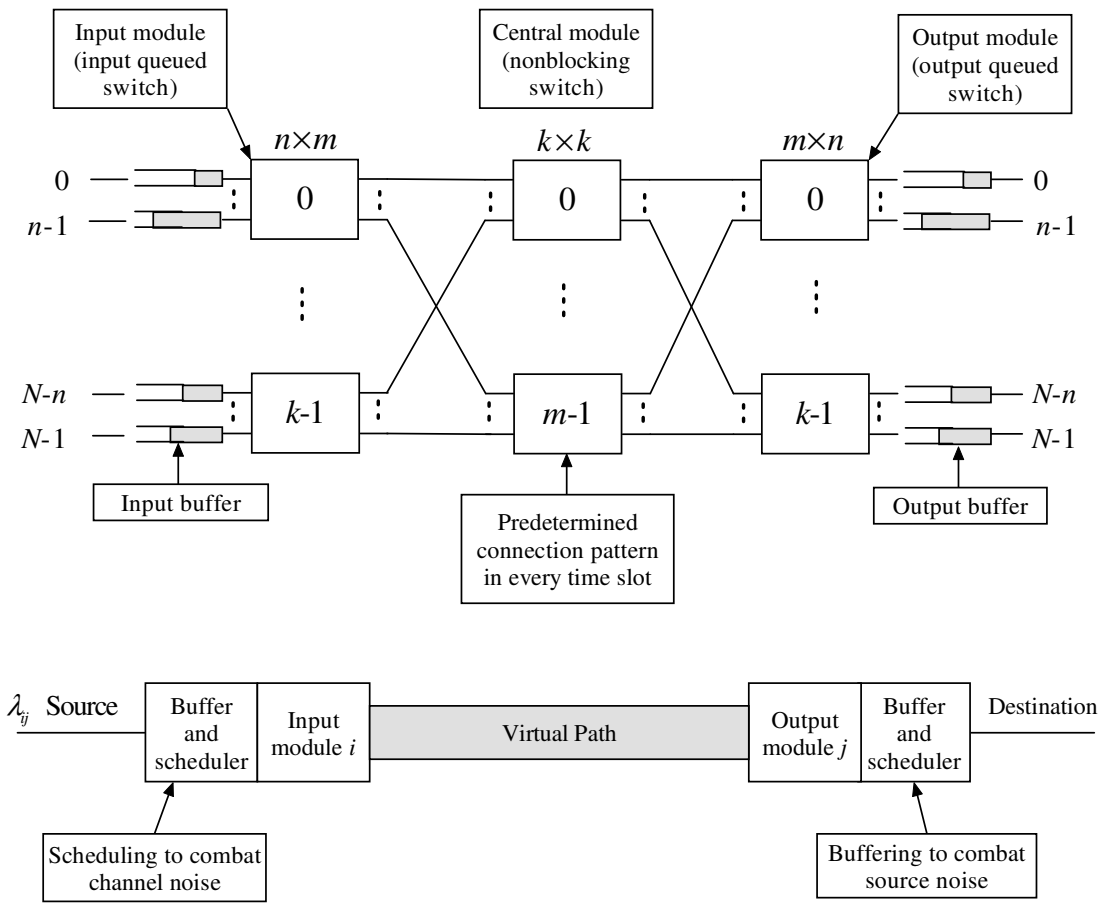

Fig. 20. Noiseless virtual path of Clos network

For any given traffic matrix $T = [\lambda_{ij}]$, where $\lambda_{ij}$ is the number of packets per time slot from input module $I_i$ to output module $O_j$, such that $\sum_i \lambda_{ij} < n \leq m$, and $\sum_j \lambda_{ij} < n \leq m$, if there exists a finite number $F$ of regular bipartite graphs such that the capacity $c_{ij}$ of the virtual path $V_{ij}$ between $I_i$ and $O_j$ satisfies

$$c_{ij} = \frac{\sum_{t=0}^{F-1} e_{ij}(t)}{F} > \lambda_{ij}, \tag{22}$$

where $e_{ij}(t)$ is the number of edges from node $i$ to node $j$ in the $t^{\text{th}}$ bipartite graph. The bandwidth requirement $T = [\lambda_{ij}]$ can be satisfied if the system periodically provides connections according to the edge coloring of these $F$ bipartite graphs. Adopting convention in the TDMA system, each cycle is called a *frame* and the period $F$ *frame size*. The routing information can be stored in the local memory of each input module to avoid the slot-by-slot computation of route assignments [36], [37].

Since the connection patterns of switch modules in the middle stage are fixed, the virtual path between any pair of input module and output module is contention-free. As shown in Fig. 20, input/output modules are scaled to a much smaller size and routing decisions in these modules become independent. Each input module is an independent input-queued switch, arrival packets can be scheduled in the input buffer according to predetermined routes in every time slot by some matching algorithm similar to the scheduling algorithm for input-queued crossbar switch [38], while each output module could be an output-queued switch similar to a knockout switch [39]. Consider the scheduled Clos network as a noiseless channel, the scheduling of path switching maps central modules and time slots into incoming packets, a process similar to the source coding of transmission [1], [2], [6], [7] if the set of predetermined connection patterns is regarded as a code book. In this section, we address the capacity allocation and traffic matrix decomposition issues, and the smoothness of scheduling will be discussed in the next section.

### *A. Capacity Allocation*

The capacity allocation problem seeks to find the capacity $c_{ij} > \lambda_{ij}$ for each virtual path $V_{ij}$ between $I_i$ and $O_j$ such that non-overbooking condition $\sum_i c_{ij} = \sum_j c_{ij} = m$ for each input/output module is observed. The problem can be formulated as constrained optimization of some objective function [40]. The choice of the objective function depends on the stochastic characteristic of the traffic on virtual paths and the quality of service requirements of connections. Following Kleinrock's independency assumption [41], [42], each virtual path can be modelled as an $M/M/1$ queue with arrival rate $\lambda_{ij}$ and service rate $c_{ij}$ for all $i, j$, then the average delay for the packets from input module $I_i$ to output module $O_j$ is given by

$$D_{ij} = \frac{1}{c_{ij} - \lambda_{ij}} \tag{23}$$

and the objective is to minimize the total weighted delay [41]

$$\sum_{i,j} \frac{\lambda_{ij}}{c_{ij} - \lambda_{ij}} \tag{24}$$

subject to $c_{ij} > \lambda_{ij}$ and $\sum_i c_{ij} = \sum_j c_{ij} = m$. The computation of the optimal allocation is quite involved in general. Nevertheless, the following heuristic algorithm based on Kleinrock's square root rule [41] can always yield suboptimal capacity allocation that is close to the optimal solution.

***Theorem** 7:* For any given traffic matrix $T = [\lambda_{ij}]$ with non-zero entries, $\lambda_{ij} > 0$ for all $i$ and $j$, then

$$\sum_{j=1}^{n} \lambda_{ij} \leq \sum_{j=1}^{n} c_{ij} = m \tag{25}$$

and

$$\sum_{i=1}^{n} \lambda_{ij} \leq \sum_{i=1}^{n} c_{ij} = m \tag{26}$$

when the capacity allocation algorithm terminates.

*Proof:* Since $\lambda_{ij} > 0$, it is obvious that

$$c_{ij}^{(k)} > c_{ij}^{(k-1)} \tag{27}$$

in $k^{th}$ iteration of the algorithm for all $k$. Also, we have

$$\sum_{j=1}^{n} c_{ij}^{(k)} \leq \sum_{j=1}^{n} c_{ij}^{(k-1)} + \sum_{j=1}^{n} \frac{A_i^{(k-1)} \sqrt{\lambda_{ij}}}{\sum_{j=1}^{n} \sqrt{\lambda_{ij}}} = m. \tag{28}$$

---

Capacity Allocation Algorithm

Step 1 Initialization: Set $k = 1$ and $c_{ij}^{(0)} = \lambda_{ij}$.

Step 2 Calculate slack capacity of each input module (row) and output module (column).

$$\begin{cases} A_i^{(k-1)} = m - \sum_{j=1}^{n} c_{ij}^{(k-1)} \\ B_j^{(k-1)} = m - \sum_{i=1}^{n} c_{ij}^{(k-1)} \end{cases}$$

Step 3 Remove any saturated row $i$ if $A_i^{(k-1)} = 0$, and column $j$ if $B_j^{(k-1)} = 0$. The algorithm terminates if there is no slack capacity left, otherwise do

$$c_{ij}^{(k)} = c_{ij}^{(k-1)} + \min\left(\frac{A_i^{(k-1)} \sqrt{\lambda_{ij}}}{\sum_{j=1}^{n} \sqrt{\lambda_{ij}}}, \frac{B_j^{(k-1)} \sqrt{\lambda_{ij}}}{\sum_{i=1}^{n} \sqrt{\lambda_{ij}}}\right),$$

and set $k = k + 1$, goto step 2.

---

The combination of (27) and (28) assures that $\sum_{j=1}^{n} c_{ij} = m$, and similarly, that $\sum_{i=1}^{n} c_{ij} = m$, when the algorithm terminates. □

Notice that if some entries $\lambda_{ij}$ are zero, then the strict inequality (27) does not hold in general, and there is no guarantee that the algorithm will halt. It is also true for any capacity allocation algorithms that if $\lambda_{ij} = 0$ implies $c_{ij} = 0$, then the doubly stochastic matrix $C = [c_{ij}]$ that satisfies (25) and (26) may not exist. Nevertheless, this problem can be eliminated in practice if a small amount of capacity $c_{ij} = \epsilon$ allocated to a virtual path with $\lambda_{ij} = 0$ is allowed.

### *B. Capacity Matrix Decomposition*

The time axis is divided into frames of time-slots, and the frame size is denoted by $F$. Within each frame, the rate requirement will be satisfied by a set of connection patterns, which are predetermined by the decomposition of the capacity matrix $[c_{ij}]$. In a time-slotted system, it is reasonable to assume that all entries $c_{ij}$ are rational numbers with a least common denominator $F$. Then $[F \cdot c_{ij}]$ is an integer doubly stochastic matrix with constant $mF$ row sum and column sum. According to theorem 4, the bipartite graph corresponding to the matrix $[F \cdot c_{ij}]$ can be colored by $mF$ colors. The colored bipartite graph can also be expressed by the genralized Birkhoff-von Neumann decomposition [43]–[45] stated in the following theorem.

***Theorem** 8: (Generalized Birkhoff-von Neumann Decomposition)* The capacity matrix $C$ has the following expansions

$$C = \frac{1}{F} \sum_{i=1}^{mF} M_i = \frac{1}{F} \sum_{i=1}^{F} G_i = \sum_{i=1}^{K} \phi_i P_i, \tag{29}$$

such that $G_i = \sum_{j=m(i-1)+1}^{mi} M_j, i = 1, \cdots, F$ and $\sum_{i=1}^{K} \phi_i = 1$.

With respect to the scheduling of path switching, the correspondence between connection patterns in the Clos network

$$[c_{ij}] = \begin{bmatrix} 1.4 & 0.5 & 1.1 \\ 0.6 & 1.8 & 0.6 \\ 1.0 & 0.7 & 1.3 \end{bmatrix} \Longrightarrow \quad F=4 \quad [e_{ij}] = \begin{bmatrix} 5 & 2 & 4 \\ 2 & 7 & 2 \\ 4 & 2 & 5 \end{bmatrix}$$

capacity matrix
(per slot)

R.M.S. error = 0.104

$$F=8 \quad [e_{ij}] = \begin{bmatrix} 11 & 4 & 8 \\ 4 & 14 & 4 \\ 8 & 5 & 10 \end{bmatrix}$$

R.M.S. error = 0.0608

Fig. 21. Round-off errors with different frame size $F$.

$C(m,n,k)$ and these matrices in the series expansions (29) is given as follows:

1) Each $M_i$ is a permutation matrix, or complete matching.
2) Each matrix $G_i$ is a sum of $m$ permutation matrices, $M_{m(i-1)+1}, \cdots, M_{mi}$, which represent connection patterns of those $m$ central modules of $C(m,n,k)$ in time slot $i$ of every frame. That is, the matrix $G_i$ is an edge colored regular bipartite graph with degree $m$. The combination of permutation matrices in $G_i$ can be arbitrary.
3) Each matrix $P_i$ is a *state* of the switch such that $P_i \neq P_j$ for all $i \neq j$, and $\{P_1, \cdots, P_K\} = \{G_1, \cdots, G_F\}$. The coefficient $\phi_i$ is the frequency of the state $P_i$ within each time frame. Since the total number of constraints in the doubly stochastic matrix $C$ equals to $(N-1)^2+1$, the number of states is bounded by $K \leq \min\{F, N^2-2N+2\}$.

The scheduling information stored in the memory is linearly proportional to $F$, the frame size is limited by the access speed and the memory space of input modules. In the capacity matrix decomposition (29), both $K$ and $F$ could be too large in practice, because the number of states $K$ is in the order of $O(N^2)$ and the frame size $F$ is determined by the least common denominator of $c_{ij}$. The following limitations compromise between complexity and QoS:

1) If the capacity matrix $C$ is bandlimited such that $c_{ij} \leq \frac{B}{F}$, then it is easy to show that $K \leq F \leq \frac{BN}{m}$.
2) The frame size $F$ can be a constant independent of switch size $N$ if $[F \cdot c_{ij}]$ is rounded off into an integer matrix, and the round-off error in the order of $O(1/F)$ is acceptable. The error can be arbitrarily small if the frame size $F$ is sufficiently large, as shown in the examples given in Fig. 21.

Consider the traffic matrix as the aggregate of signals input to the packet switch, the series expansion of doubly stochastic capacity matrix into permutation matrices (29), and the reconstruction of capacity by weighted running sum of scheduled connections are mathematically similar to that of the Nyquist-Shannon sampling theorem of bandlimited signals [18]–[21] input to the transmission channel.

***Sampling theorem of bandlimited signal*** *If a function $f(t)$ contains no frequencies higher than $W$ cps, it is completely determined by its ordinates at a series of equally spaced sampling intervals of $1/2W$ sec., called Nyquist intervals. If $|F(\omega)| = 0$, for $|\omega| \geq 2\pi W$, then*

$$f(t) = \sum_{-\infty}^{+\infty} f_n \frac{sin\pi(2Wt-n)}{\pi(2Wt-n)}$$

*where $f_n = f(\frac{n}{2W})$ is the $n^{th}$ sample of $f(t)$.*

An important common characteristic of both series expansions is the complexity reduction of communication systems. Without scheduling, the total number of possible permutations of a packet switch for a given capacity matrix is $N!$. The complexity reduction is achieved by limiting the space to those permutations only involved in the expansion (29). Hence, the dimension of permutation space is reduced to $O(N^2)$ by the traffic matrix decomposition. It can further be reduced to $O(N)$ if each connection is bandwidth limited. The complexity of packet level problems can be significantly reduced if the switch fabric is limited to this set of predetermined permutations only, operating in a much smaller subspace, yet the capacity of each connection is still guaranteed. In general, the minimal number of permutations required is determined by the edge coloring of the corresponding bipartite graphs as stated in theorem 4.

In digital transmission, the sampling theorem reduces the infinite dimensional signal space in any time interval $T$ to a finite number $2TW$ of samples, and the minimal number of samples is determined by the Nyquist sampling rate for limited bandwidth $W$ of the input signal function. The comparisons of the these two expansions and their respective roles in packet switching and digital transmission are listed in Table III.

## VI. Scheduling and Source Coding

The process and function of a scheduling algorithm are similar to those of source coding in transmission. If we regard the set of predetermined connection patterns as a code book, scheduling incoming packets in the input buffer according to predetermined connection patterns is the same as encoding the source signal. This point will be elaborated by the construction of a scheduling algorithm based on the Huffman tree. We first show that the smoothness of scheduling, like source coding, is bounded by the source entropy, or the entropy of capacity decomposition. The results can then be generalized to the smoothness of two-dimensional scheduling of tokens assigned to each virtual path of the Clos network.

### A. Smoothness of Scheduling

In light of the fairness of services, the scheduling is expected to be as smooth as possible. In a Clos network with path switching, in addition to guarantee capacity for each virtual path, smooth scheduling also reduce delay jitters and alleviate *head-of-line* (HOL) blocking at input buffer [25].

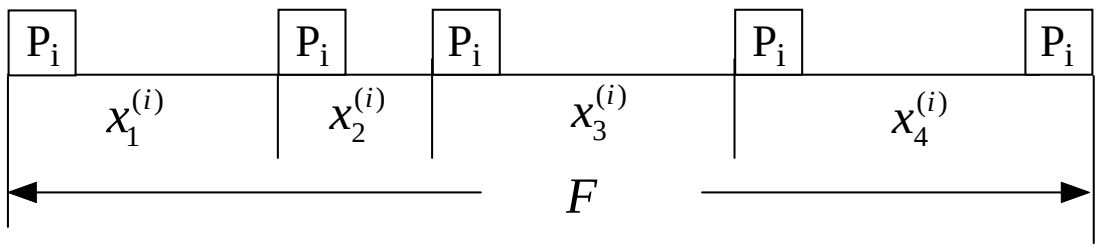

Fig. 22. Interstate time for $i^{th}$ state within a frame of size $F$

| | Packet switching | Digital transmission |
|---|---|---|
| Network environment | Time-slotted switching system | Time-slotted transmission system |
| Bandwidth limitation | Capacity limited traffic matrix<br>$\sum_{i=1}^{N}\lambda_{ij} \leq \sum_{i=1}^{N} c_{ij} = m$<br>$\sum_{j=1}^{N}\lambda_{ij} \leq \sum_{i=1}^{N} c_{ij} = m$<br>$\lambda_{ij} \leq c_{ij}$ | Bandwidth limited signal function<br>$f(t) = \frac{1}{2\pi}\int_{-2\pi w}^{2\pi w} F(\omega)d\omega$<br>$F(\omega) = 0, \qquad \lvert\omega\rvert \geq 2\pi w$ |
| Samples | $(0, 1)$ Permutation matrices | $(0, 1)$ Binary sequences |
| Expansion | BvN decomposition (Hall's marriage theorem)<br>$C = \sum_{i=1}^{Fm}\frac{M_i}{F} = \sum_{i=1}^{K}\phi_i P_i$<br>Frame size=$F$=lcm of denominators of $c_{ij}$ | Fourier series<br>$f(t) = \sum_{-\infty}^{+\infty} f_n \frac{sin\pi(2Wt-n)}{\pi(2Wt-n)}$<br>Nyquist interval= $T = \frac{1}{2W}$ |
| Inversion | Reconstruct the capacity by running sum | Reconstruct the signal by interpolation |
| Complexity reduction | • Reduce number of permutations from $N!$ to $O(N^2)$<br>• Reduce to $O(N)$ if bandwidth is limited<br>• Reduce to constant $F$ if truncation error of order $O(1/F)$ is acceptable | • Reduce infinite dimensional signal space to finite number $2tW$ in any duration $t$ |
| QoS | Capacity guarantee, Scheduling, Delay bound | Error-correcting code, Data compression, DSP |

TABLE III
COMPARISON OF TRAFFIC MATRIX DECOMPOSITION AND SAMPLING THEOREM

Consider a scheduling of a given capacity decomposition $C = \sum_{i=1}^{K} \phi_i P_i$ with frame size $F$. Let $X_1^{(i)}, X_2^{(i)}, \cdots, X_{n_i}^{(i)}$ be a sequence of interstate times of state $P_i$ within a frame. It is easy to see from Fig. 22 that

$$n_i = \phi_i F; \tag{30}$$

$$X_1^{(i)} + \cdots + X_{n_i}^{(i)} = F, \qquad \text{for all } i = 1, \cdots, k. \tag{31}$$

Since delay jitter is one of the major concerns of scheduling, it is natural to define the smoothness of scheduling by the second moment of interstate time.

***Definition** 5:* The smoothness of state $P_i$ is defined by

$$L_i = \log\sqrt{\frac{\sum_{k=1}^{n_i}(X_k^{(i)})^2}{n_i}}, \tag{32}$$

and the smoothness of a scheduling is the weighted average given by

$$L = \sum_{i=1}^{K}\phi_i L_i. \tag{33}$$

The following theorem reveals the fact that the smoothness of scheduling is parallel to the code length of source coding in information theory.

***Theorem** 9:* For any scheduling of a given capacity decomposition $C = \sum_{i=1}^{K}\phi_i P_i$ with frame size $F$, we have

$$\sum_{i=1}^{K} 2^{-L_i} \leq 1 \qquad \text{(Kraft's inequality)} \tag{34}$$

and the average smoothness is bounded by

$$L = \sum_{i=1}^{K}\phi_i L_i \geq \sum_{i=1}^{K}\phi_i \log\frac{1}{\phi_i} = H. \tag{35}$$

Both equalities hold when $X_k^{(i)} = \frac{1}{\phi_i}$, for all $i = 1, \cdots, K$, and $k = 1, \cdots, n_i$.

*Proof:* The inequality (34) can be obtained by substituting (30) and (31) into the Cauchy-Schwartz inequality

$$\sum_{i=1}^{K} 2^{-L_i} = \sum_{i=1}^{K}\sqrt{\frac{n_i}{\sum_{k=1}^{n_i}(X_k^{(i)})^2}} \leq \sum_{i=1}^{K}\frac{n_i}{F} = \sum_{i=1}^{K}\phi_i = 1.$$

Again, using Cauchy-Schwartz inequality, we have

$$\begin{aligned} L &= \sum_{i=1}^{K}\phi_i L_i = \sum_{i=1}^{K}\phi_i \log\sqrt{\frac{\sum_{k=1}^{n_i}(X_k^{(i)})^2}{n_i}} \\ &= \sum_{i=1}^{K}\frac{\phi_i}{2}\log\frac{\sum_{k=1}^{n_i}(X_k^{(i)})^2}{n_i} \geq \sum_{i=1}^{K}\phi_i \log\frac{1}{\phi_i} = H. \end{aligned}$$

□

Suppose equalities in the above theorem hold, then the scheduling is obviously optimal. First, we consider a simple example when $K = F, \phi_i = \frac{1}{F}$, and $n_i = 1$ for all $i$, then $X_1^{(i)} = F$ for all $i = 1, \cdots, F$, in which case the smoothness equals entropy.

$$L = \frac{1}{F}\sum_{i=1}^{F}\log\sqrt{(F^2)} = \log F = H \tag{36}$$

When all weights are reciprocals of the power of 2, for example $\phi_i = \frac{1}{2}, \frac{1}{4}, \frac{1}{8}, \frac{1}{8}$, an optimal scheduling can be easily found. An example is shown in Fig 23(a), where $F = 8$, $K = 4$, and $n_i = 4, 2, 1, 1$ with respect to state $P_i$ for $i = 1, \cdots, 4$.

| | | | | | | | |
|---|---|---|---|---|---|---|---|
| $P_1$ | $P_2$ | $P_1$ | $P_3$ | $P_1$ | $P_2$ | $P_1$ | $P_4$ |

(a) Optimal Scheduling

| | | | | | | | |
|---|---|---|---|---|---|---|---|
| $P_1$ | $P_1$ | $P_2$ | $P_1$ | $P_1$ | $P_2$ | $P_3$ | $P_4$ |

(b) *WFQ* Scheduling

Fig. 23. Examples of Scheduling

The scheduling with constant interstate time $X^{(i)} = 2, 4, 8, 8$ satisfies both equalities of theorem 10

$$H = L = \frac{1}{2} \cdot + \frac{1}{4} \cdot 2 + \frac{1}{8} \cdot 3 + \frac{1}{8} \cdot 3 = 1.75. \quad (37)$$

Another scheduling based on weighted fair queueing (*WFQ*) for the same decomposition is shown in Fig 23(b). The smoothness $L = 1.8758$ is greater than the decomposition entropy $H = 1.75$, and the superiority of the optimal scheduling is quite obvious by comparing Fig. 23(a) and (b). An upper bound of smoothness is given in the following theorem.

***Theorem** 10:* For any capacity decomposition $C = \sum_{i=1}^{K} \phi_i P_i$, it is always possible to device a scheduling whose smoothness $L$ is within $\frac{1}{2}$ of decomposition entropy

$$H \leq L < H + \frac{1}{2} \quad (38)$$

*Proof:* Consider a random scheduling without frame structure such that the state $P_i$ of each time slot is randomly selected with probability $\phi_i$. The interstate time $X^{(i)}$ of state $P_i$ is a geometric random variable with $E[X^{(i)}] = \frac{1}{\phi_i}$, and $Var[X^{(i)}] = \frac{1-\phi_i}{\phi_i^2}$. The second moment of $X^{(i)}$ and the smoothness of state $P_i$ are given respectively as follows

$$E[(X^{(i)})^2] = Var[X^{(i)}] + (E[X^{(i)}])^2 = \frac{2 - \phi_i}{\phi_i^2}, \quad (39)$$

and

$$L_i = \log \sqrt{E[(X^{(i)})^2]} = \frac{1}{2} \log (2 - \phi_i) + \log \phi_i. \quad (40)$$

It is counter intuitive that the smoothness $L$ of random scheduling does not equal to the decomposition entropy $H$; in fact we have

$$L = \sum_{i=1}^{K} \phi_i L_i = \frac{1}{2} \sum_{i=1}^{K} \phi_i \log (2 - \phi_i) + H. \quad (41)$$

It is easy to show that the Kullback-Leibler distance $L - H$ reaches the maximum value when $\phi_1 = \phi_2 = \cdots = \phi_K = \frac{1}{K}$, in which case we have

$$L - H = \frac{1}{2} \sum_{i=1}^{K} \phi_i \log (2 - \phi_i) \leq \frac{1}{2} \log (2 - \frac{1}{K}) < \frac{1}{2}. \quad (42)$$

The random scheduling is actually no scheduling at all. Hence, there should exist a scheduling algorithm, with or without frame, that satisfies the bound given in (38). □

It is obvious that theorem 9 and 10 are counterparts of the source coding theorem stated as follows.

***Noiseless coding theorem*** *Let random variable $X$ take the possible values $x_1, \cdots, x_K$ with respective probabilities $p(x_1), \cdots, p(x_K)$. Then, the necessary and sufficient condition to encode the values of $X$ in binary prefix code ( none of which is an extension of another) of respective lengths $L_1, \cdots, L_K$ is*

$$\sum_{i=1}^{K} 2^{-L_i} \leq 1 \qquad \textit{(Kraft's inequality)} \quad (43)$$

*The average code length is bounded by*

$$L = \sum_{i=1}^{K} p(x_i) L_i \geq H(X) = \sum_{i=1}^{K} p(x_i) \log \frac{1}{p(x_i)} \quad (44)$$

*and it is always possible to device an optimal prefix code for $X$ whose average code length $L$ is within 1 of entropy*

$$H(X) \leq L < H(X) + 1. \quad (45)$$

### B. Comparison of Scheduling Algorithms

The scheduling algorithm for an arbitrary set of decomposition weights $\phi_i$ achieves optimal smoothness if $L = H$, which requires the constant interstate time that equals to the inverse of weight $X_k^{(i)} = \frac{F}{n_i} = \frac{1}{\phi_i}$ for all state $P_i$. Most proposed scheduling algorithms are indeed constructed from the reciprocal of the normalized weights. A comparison of several stereotype algorithms is given below.

*1) Weighted Fair Queueing (WFQ) Scheduling Algorithm:* The *WFQ* scheduling algorithm [46] is constructed from the sequence of virtual finish times $\frac{1}{\phi_i}, \frac{2}{\phi_i} \cdots$ of each state $P_i$ with the initial finish time $\frac{1}{\phi_i}$. The algorithm is given as follows.

---

The *WFQ* scheduling algorithm:
Select the state with the smallest finish time and increase its finish time by the inverse of its weight. Repeat this process until the frame size is reached

---

We will use the running example of a set of five states $P_1, P_2, P_3, P_4, P_5$ with respective weights $\phi_1 = 0.5, \phi_2 = \phi_3 = \phi_4 = \phi_5 = 0.125$ to illustrate the performance of those scheduling algorithms presented in this section. The sequence generated by *WFQ* in each time slot $\tau$ within a frame is given in the Table IV, and the final sequence is $P_1P_1P_1P_1P_2P_3P_4P_5$.

| $\tau$ | $P_1$ | $P_2$ | $P_3$ | $P_4$ | $P_5$ | Selection |
|---|---|---|---|---|---|---|
| 1 | 2 | 8 | 8 | 8 | 8 | $P_1$ |
| 2 | 4 | 8 | 8 | 8 | 8 | $P_1$ |
| 3 | 6 | 8 | 8 | 8 | 8 | $P_1$ |
| 4 | 8 | 8 | 8 | 8 | 8 | $P_1$ |
| 5 | 10 | 8 | 8 | 8 | 8 | $P_2$ |
| 6 | 10 | 16 | 8 | 8 | 8 | $P_3$ |
| 7 | 10 | 16 | 16 | 8 | 8 | $P_4$ |
| 8 | 10 | 16 | 16 | 16 | 8 | $P_5$ |

TABLE IV
AN EXAMPLE OF $WFQ$ ALGORITHM

Although the *WFQ* algorithm properly utilizes the inverse weights in constructing the scheduling, the above example shows that the sequence generated by *WFQ* is still far from the achievable optimal smoothness. Possible amendments to *WFQ* are discussed below.

*2) $WF^2Q$ Scheduling Algorithm:* $WF^2Q$ is a scheduling algorithm [47] that incorporates the stringent rate requirement of *generalized processor sharing (GPS)* [48] in *WFQ*. Let $T_i(\tau)$ be the number of time slots assigned to state $P_i$ up to time $\tau$, the $WF^2Q$ will select the state $P_i$ that satisfies

$$T_i(\tau - 1) < \tau \cdot \phi_i \quad (46)$$

in every time slot $\tau = 1, 2, \cdots$. That is, only the states that have started its service in the corresponding *GPS* system will be selected. The set of states qualified for selection in time slot $\tau$ is defined by

$$Q_\tau = \{P_i \mid T_i(\tau - 1) < \tau \cdot \phi_i\} \quad (47)$$

Hence, all states are qualified in the first time slot (see Table IV), because $T_i(0) = 0$ for all $i$. The set of qualified states $Q_1 = \{P_1P_2P_3P_4P_5\}$ is then ordered by their finish time and the state $P_1$ is selected. In the second time slot, however, $T_1(1) = 1$ and $2 \cdot \phi_1 = 1$, so the state $P_1$ cannot be selected and $Q_2 = \{P_2P_3P_4P_5\}$. In each time slot, whenever a state is selected, like in $WFQ$, its finish time will be increased by the inverse of its weight.

The following table provides the whole procedure of $WF^2Q$. The final sequence is $P_1P_2P_1P_3P_1P_4P_1P_5$, which is better than the sequence produced by $WFQ$. Another approach to amend *WFQ* based on the Huffman tree is given next.

| $\tau$ | $P_1$ | $P_2$ | $P_3$ | $P_4$ | $P_5$ | $Q_\tau$ | Selection |
|---|---|---|---|---|---|---|---|
| 1 | 2 | 8 | 8 | 8 | 8 | $P_1P_2P_3P_4P_5$ | $P_1$ |
| 2 | 4 | 8 | 8 | 8 | 8 | $P_2P_3P_4P_5$ | $P_2$ |
| 3 | 4 | 8 | 8 | 8 | 8 | $P_1P_3P_4P_5$ | $P_1$ |
| 4 | 4 | 16 | 8 | 8 | 8 | $P_3P_4P_5$ | $P_3$ |
| 5 | 6 | 16 | 8 | 8 | 8 | $P_1P_4P_5$ | $P_1$ |
| 6 | 6 | 16 | 16 | 8 | 8 | $P_4P_5$ | $P_4$ |
| 7 | 8 | 16 | 16 | 8 | 8 | $P_1P_5$ | $P_1$ |
| 8 | 10 | 16 | 16 | 16 | 8 | $P_5$ | $P_5$ |

TABLE V
AN EXAMPLE OF $WF^2Q$ ALGORITHM

*3) Huffman Round Rabin (HuRR) Algorithm :* It is known that the Huffman code [49] is the optimal source code constructed from the binary probability tree, called the *Huffman tree*, in a hierarchical manner. When the *WFQ* is applied to only two states, the average interstate time of each state is very close to the optimum because the two-state Huffman binary tree only has one level.

For example, consider the two states $P_1$ and $P_2$ with respective weights $\phi_1 = 0.25$ and $\phi_2 = 0.75$, the *WFQ* will produce the sequence $P_2P_2P_1P_2$. The interstate time of state $P_1$ is 4, which is exactly the reciprocal of 0.25, and state $P_2$ has interstate time of 1 and 2, which are the two integers nearest $1/0.75$. However, if there are more than two states, the *WFQ* fails to achieve the optimality because the Huffman tree has multiple levels.

The optimality of Huffman code is closely related to Kraft's inequality, which is the necessary and sufficient condition of prefix source coding. In scheduling, the optimal smoothness $L = H$ also implies $L_i = \log \frac{1}{\phi_i}$ for all state $P_i$. Hence, the equality of Kraft's inequality holds

$$\sum_{i=1}^{K} 2^{-L_i} = 1$$

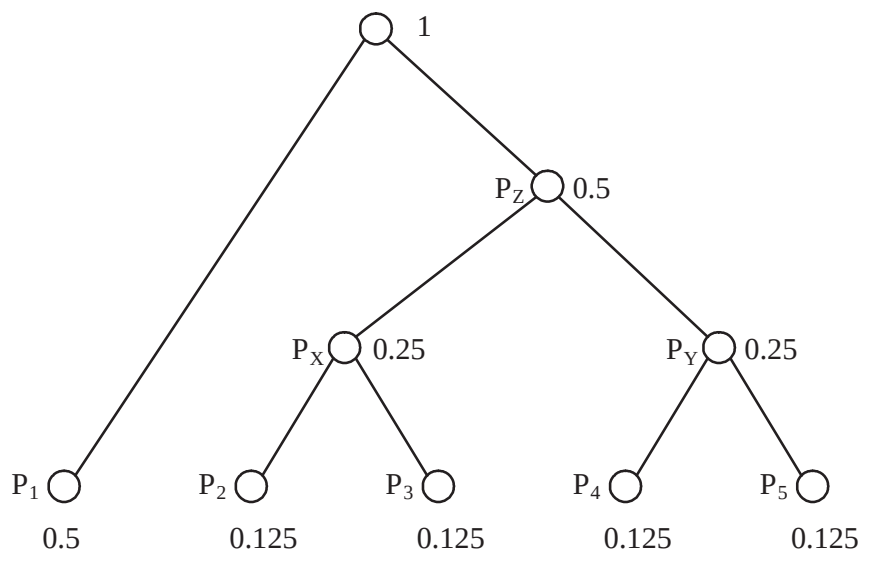

Fig. 24. A Huffman tree of HuRR Algorithm

Similar to the Huffman code, the *HuRR* scheduling algorithm proposed in [50] is a hierarchical scheduling algorithm synthesized from the Huffman tree.

Consider each state of the decomposition as a symbol and the weight of the state as its probability, a Huffman tree can be constructed from ordered sequence of probabilities as usual. The probability of the root node is 1, the probability of each leave node is the probability of the symbol represented by that leave, and the probability of each intermediate node is the sum of the probabilities of two successors. The *HuRR* algorithm comprises the following steps.

The *HuRR* scheduling algorithm:

- Step 1 Initially set the root to be temporary node $P_X$, and $S = P_X \cdots P_X$ to be a temporary sequence.
- Step 2 Apply the *WFQ* to the two successors of $P_X$ to produce a sequence $T$, and substitute $T$ for the subsequence $P_X \cdots P_X$ of $S$.
- Step 3 If there is no intermediate node in the sequence $S$, then terminate the algorithm. Otherwise select an intermediate node $P_X$ appearing in $S$ and goto step 2.

The Huffman tree of the previous example is shown in Fig. 24, the HuRR scheduling algorithm will generate the following sequences

$$\begin{aligned} P_1P_ZP_1P_ZP_1P_ZP_1P_Z &\rightarrow P_1P_XP_1P_YP_1P_XP_1P_Y \\ &\rightarrow P_1P_2P_1P_4P_1P_3P_1P_5 \end{aligned}$$

In this particular example, the following Huffman code can be generated from the tree depicted in Fig.(24)

$$P_1 \longleftarrow 0, P_2 \longleftarrow 100, P_3 \longleftarrow 101, P_4 \longleftarrow 110, P_5 \longleftarrow 111$$

and the logarithm of interstate time of each state is equal to the length of its Huffman code. The smoothness of the sequence of this example is the same as that generated by $WF^2Q$ earlier. However, the *HuRR* outperforms $WF^2Q$ in general as demonstrated by the comparison given next.

*4) Comparison of Smoothness of Scheduling Algorithms:* The smoothness of preceding scheduling algorithms are compared with random scheduling in Table VI to show their progressive improvements. It is clear that the sequences generated by *HuRR* are, in general, closer to the entropy than the other two. The performance of $WF^2Q$ is comparable to *HuRR*, while *WFQ* is not satisfactory in some cases.

| $P_1$ | $P_2$ | $P_3$ | $P_4$ | *Random* | *WFQ* | *WF*$^2$*Q* | *HuRR* | *Entropy* |
|---|---|---|---|---|---|---|---|---|
| 0.1 | 0.1 | 0.1 | 0.7 | 1.628 | 1.575 | 1.414 | 1.414 | 1.357 |
| 0.1 | 0.1 | 0.2 | 0.6 | 1.894 | 1.734 | 1.626 | 1.604 | 1.571 |
| 0.1 | 0.1 | 0.3 | 0.5 | 2.040 | 1.784 | 1.724 | 1.702 | 1.686 |
| 0.1 | 0.2 | 0.2 | 0.5 | 2.123 | 1.882 | 1.801 | 1.772 | 1.761 |
| 0.1 | 0.1 | 0.4 | 0.4 | 2.086 | 1.787 | 1.745 | 1.745 | 1.722 |
| 0.1 | 0.2 | 0.3 | 0.4 | 2.229 | 1.903 | 1.903 | 1.884 | 1.847 |
| 0.2 | 0.2 | 0.2 | 0.4 | 2.312 | 2.011 | 1.980 | 1.933 | 1.922 |
| 0.1 | 0.3 | 0.3 | 0.3 | 2.286 | 1.908 | 1.908 | 1.908 | 1.896 |
| 0.2 | 0.2 | 0.3 | 0.3 | 2.370 | 2.016 | 2.016 | 1.980 | 1.971 |

TABLE VI
THE SMOOTHNESS COMPARISON OF SCHEDULING ALGORITHMS

The significance of the Huffman coding scheme lies in the structure of the Huffman Tree. By the same token, *HuRR*, which is also implemented with the Huffman tree, can provide the best performance in terms of smoothness among the scheduling algorithms discussed in this section. However, *HuRR* has not yet been proven to be the optimal scheduling algorithm. It is clear that the key to construct optimal scheduling is to explore the structure of the smoothness $L_i$ of each state $P_i$ in a frame of size $F$ and Kraft's inequality that is satisfied by $L_i$, which is mathematically equivalent to the codeword length in source coding theorem.

### *C. Two-dimensional Scheduling*

The smoothness of scheduling described in the preceding section is for a single-server system. However, the capacity matrix decomposition of the packet switch system is a two-dimensional scheduling aimed at multiple inputs and outputs [51]. The generalization of the smoothness to two-dimensional scheduling is discussed in this section.

Recall that the capacity matrix $C = [c_{ij}]$ of a $C(m, n, k)$ Clos network is a $k \times k$ matrix, where $c_{ij}$ is the average number of tokens assigned to the virtual path $V_{ij}$ between input module $i$ and output module $j$. If the frame size is $F$, then the matrix $F \cdot C$ is integer doubly stochastic, such that $\sum_i^k c_{ij} = m, \sum_j^k c_{ij} = m$.

In the following discussion of two-dimensional smoothness, we will consider, without loss of generality, capacity matrix $C = [c_{ij}]$ and $\sum_i^N c_{ij} = 1, \sum_j^N c_{ij} = 1$ of a crossbar switch for the sake of simplicity. The two-dimensional entropies of the capacity matrix $C$ are defined as follows.

***Definition** 6:* Let $C = [c_{ij}]$ be a doubly stochastic capacity matrix; the entropy of input module $i$ is defined by

$$\overline{H}_i = -\sum_{j=1}^{N} c_{ij} \log c_{ij}, \tag{48}$$

and the the set of entropies of input modules is represented by the column vector

$$\overline{H} = \begin{bmatrix} \overline{H}_1 \\ \overline{H}_2 \\ \vdots \\ \overline{H}_N \end{bmatrix} = \left[\overline{H}_1, \overline{H}_2, \cdots, \overline{H}_N\right]^T. \tag{49}$$

Similarly, the entropy of output module $j$ is defined by

$$\underline{H}_j = -\sum_{i=1}^{N} c_{ij} \log c_{ij}, \tag{50}$$

and the set of entropies of output modules is represented by the row vector

$$\underline{H} = [\underline{H}_1, \underline{H}_2, \cdots, \underline{H}_N]. \tag{51}$$

The entropy of the capacity matrix $C$ is defined by

$$H(C) = \sum_{i=1}^{N} \overline{H}_i = \sum_{j=1}^{N} \underline{H}_j = -\sum_{j=1}^{N}\sum_{i=1}^{N} c_{ij} \log c_{ij}. \tag{52}$$

The smoothness of a two-dimensional scheduling is defined by the inter-token time in the same manner as the interstate time of single server scheduling. Let $n_{ij}$ be the number of tokens assigned to the virtual path $V_{ij}$ within a frame $F$, and $y_1^{(ij)}, y_2^{(ij)}, \cdots, y_{n_{ij}}^{(ij)}$ be the sequence of the inter-token times. We have

$$n_{ij} = c_{ij} F; \tag{53}$$

$$y_1^{(ij)} + y_2^{(ij)} + \cdots + y_{n_{ij}}^{(ij)} = F \tag{54}$$

for all $i, j = 1, \cdots, N$. It should be noted that the number of tokens assign to a virtual path $V_{ij}$ in a time slot can be more than one if the number of central modules $m > 1$, in which case the degenerate inter-token time $y_k^{(ij)} = 0$ will be allowed. The smoothness of two-dimensional scheduling is defined as follows.

***Definition** 7:* For a given scheduling of the capacity matrix decomposition $C = \sum_{i=1}^K \phi_i P_i$, the smoothness of the virtual path $V_{ij}$ is defined by

$$d_{ij} = \log \sqrt{\frac{\sum_{k=1}^{n_{ij}} \left(y_k^{(ij)}\right)^2}{n_{ij}}}. \tag{55}$$

The smoothness of input module $i$ is defined by

$$\overline{D}_i = \sum_{j=1}^{N} c_{ij} d_{ij} = \sum_{j=1}^{N} c_{ij} \log \sqrt{\frac{\sum_{k=1}^{n_{ij}} (y_k^{(ij)})^2}{n_{ij}}}, \tag{56}$$

and the input smoothness is the column vector

$$\overline{D} = \begin{bmatrix} \overline{D}_1 \\ \overline{D}_2 \\ \vdots \\ \overline{D}_N \end{bmatrix} = \left[\overline{D}_1, \overline{D}_2, \cdots, \overline{D}_N\right]^T. \tag{57}$$

Similarly, the smoothness of output module $j$ is defined by

$$\underline{D}_j = \sum_{i=1}^{N} c_{ij} d_{ij} = \sum_{i=1}^{N} c_{ij} \log \sqrt{\frac{\sum_{k=1}^{n_{ij}} (y_k^{(ij)})^2}{n_{ij}}}, \qquad (58)$$

and the output smoothness is the row vector

$$\underline{D} = [\underline{D}_1, \underline{D}_2, \cdots, \underline{D}_N] \qquad (59)$$

The smoothness of the two-dimensional scheduling is defined by

$$D = \sum_{i=1}^{N} \overline{D}_i = \sum_{j=1}^{N} \underline{D}_j = \sum_{j=1}^{N} \sum_{i=1}^{N} c_{ij} d_{ij}. \qquad (60)$$

The properties of two-dimension smoothness are similar to that of single server scheduling. The proof of the following theorem is the same as that of theorem 9.

***Theorem** 11:* For capacity matrix decomposition $C = \sum_{i=1}^{K} \phi_i P_i$ with frame size $F$, any scheduling satisfies the following smoothness inequalities

1) The matrix $K_r = \left[2^{-d_{ij}}\right]$, called *Kraft's Matrix*, is doubly sub-stochastic such that

$$\sum_{i=1}^{N} 2^{-d_{ij}} \leq 1; \qquad \sum_{j=1}^{N} 2^{-d_{ij}} \leq 1. \qquad (61)$$

2) For input module $i = 1, \cdots, N$, we have

$$\overline{D}_i \geq \overline{H}_i. \qquad (62)$$

3) For output module $j = 1, \cdots, N$, we have

$$\underline{D}_j \geq \underline{H}_j. \qquad (63)$$

4) The overall smoothness is bounded by

$$D \geq H(C). \qquad (64)$$

The above equalities hold when $y_k^{(ij)} = \frac{1}{c_{ij}}$, for all $i, j = 1, 2, \cdots, N$ and $k = 1, 2, \cdots, n_{ij}$.

The difference between smoothness and entropy can be considered as the distortion of delay jitter introduced by scheduling. For circuit switch, the capacity matrix $C$ is a permutation matrix, in which case we have $H(C) = D = 0$. Another extreme case is the scheduling of uniform capacity matrix $C = [\frac{1}{N}] = \frac{1}{N} \sum_{i=1}^{N} P_i$ with frame size $F = N$. In this case, the entropy $H(C) = N \log N$ also equals to the smoothness $D = N \log N$ because inter-token time $y_k^{(ij)} = N$, for all $i, j = 1, 2, \cdots, N$ and $k = 1, 2, \cdots, n_{ij}$ for any scheduling, in which case the entropy $H(C)$ is the maximum because the capacity matrix is completely uniform, and the smoothness of any scheduling cannot make it any worse.

It is clear that if the Kraft matrix

$$K_r = \left[2^{-d_{ij}}\right]$$

is doubly stochastic then the two-dimensional scheduling is optimal. However, the existence of the optimal scheduling that can reach the entropy $H(C)$ of any capacity matrix $C$ is still an open issue. Nevertheless, a rule of thumb is to minimize $K$, the number of permutation matrices in the decomposition [52], since the maximal delay bound is on the order of $O(K)$ for most known scheduling algorithms [19], [53].

For example, given a doubly stochastic matrix $C$,

$$C = \begin{bmatrix} 0.75 & 0 & 0.125 & 0.125 \\ 0.125 & 0.5 & 0.375 & 0 \\ 0.125 & 0.125 & 0.5 & 0.25 \\ 0 & 0.375 & 0 & 0.625 \end{bmatrix}$$

The input entropy $\overline{H}$ and output entropy $\underline{H}$ of matrix $C$ are given respectively by

$$\overline{H} = \begin{bmatrix} 1.0613 & 1.4056 & 1.7500 & 0.9544 \end{bmatrix}^T$$

and

$$\underline{H} = \begin{bmatrix} 1.0613 & 1.4056 & 1.4056 & 1.2988 \end{bmatrix}$$

The entropy of the capacity matrix $C$ is $H(C) = 5.1714$. One possible decomposition is given by

$$\begin{aligned} 8 \cdot C \quad = \quad & \begin{bmatrix} 6 & 0 & 1 & 1 \\ 1 & 4 & 3 & 0 \\ 1 & 1 & 4 & 2 \\ 0 & 3 & 0 & 5 \end{bmatrix} = 4 \begin{bmatrix} 1 & 0 & 0 & 0 \\ 0 & 1 & 0 & 0 \\ 0 & 0 & 1 & 0 \\ 0 & 0 & 0 & 1 \end{bmatrix} \\ & + \begin{bmatrix} 1 & 0 & 0 & 0 \\ 0 & 0 & 1 & 0 \\ 0 & 1 & 0 & 0 \\ 0 & 0 & 0 & 1 \end{bmatrix} + \begin{bmatrix} 1 & 0 & 0 & 0 \\ 0 & 0 & 1 & 0 \\ 0 & 0 & 0 & 1 \\ 0 & 1 & 0 & 0 \end{bmatrix} \\ & + \begin{bmatrix} 0 & 0 & 1 & 0 \\ 1 & 0 & 0 & 0 \\ 0 & 0 & 0 & 1 \\ 0 & 1 & 0 & 0 \end{bmatrix} + \begin{bmatrix} 0 & 0 & 0 & 1 \\ 0 & 0 & 1 & 0 \\ 1 & 0 & 0 & 0 \\ 0 & 1 & 0 & 0 \end{bmatrix} \end{aligned}$$

in which $K = 5$ and the permutation matrices are denoted by $\{P_1, P_2, \cdots, P_5\}$, respectively. If the *WFQ* scheduling algorithm is applied to this set of permutation matrices, the resulting sequence is

$$P_1 P_1 P_1 P_1 P_2 P_3 P_4 P_5$$

which gives rise to the following two-dimensional scheduled tokens

$$\begin{matrix} a & a & a & a & a & a & b & c \\ b & b & b & b & c & d & d & d \\ c & c & c & c & b & b & a & b \\ d & d & d & d & d & c & c & a \end{matrix}$$

where each row corresponds to an output module and the symbol $a, b, c$ and $d$ represents the respective tokens assigned to input module 1 to 4. The input smoothness of this scheduling is given by

$$\overline{D} = \begin{bmatrix} 1.2084 & 1.6997 & 2.0323 & 1.3118] \end{bmatrix}^T$$

and the output smoothness is

$$\underline{D} = \begin{bmatrix} 1.2084 & 1.7636 & 1.6997 & 1.5805 \end{bmatrix}$$

which sum up to $D = 6.2522$. When the *HuRR* scheduling algorithm is applied to the same set of permutation matrices, the resulting sequence is

$$P_1 P_2 P_1 P_4 P_1 P_3 P_1 P_5$$

which gives rise to the following two-dimensional scheduled tokens

$$\begin{array}{cccccccc} a & a & a & b & a & a & a & c \\ b & c & b & d & b & d & b & d \\ c & b & c & a & c & b & c & b \\ d & d & d & c & d & c & d & a \end{array}$$

This result is obviously smoother than the previous one as evident by the following smoothness measures

$$\overline{D} = \left[ \begin{array}{cccc} 1.1250 & 1.4375 & 1.7902 & 1.0267 \end{array} \right]^T$$

and

$$\underline{D} = \left[ \begin{array}{cccc} 1.1250 & 1.4375 & 1.4375 & 1.3794 \end{array} \right]$$

which sum up to $D = 5.3794$, an improvement consistent with the comparison of single server scheduling described in the preceding section.

Suppose we consider another capacity matrix decomposition

$$\begin{aligned} 8 \cdot C \quad &= \begin{bmatrix} 0 & 0 & 1 & 0 \\ 1 & 0 & 0 & 0 \\ 0 & 1 & 0 & 0 \\ 0 & 0 & 0 & 1 \end{bmatrix} + \begin{bmatrix} 0 & 0 & 0 & 1 \\ 0 & 0 & 1 & 0 \\ 1 & 0 & 0 & 0 \\ 0 & 1 & 0 & 0 \end{bmatrix} \\ &+2 \begin{bmatrix} 1 & 0 & 0 & 0 \\ 0 & 0 & 1 & 0 \\ 0 & 0 & 0 & 1 \\ 0 & 1 & 0 & 0 \end{bmatrix} + 4 \begin{bmatrix} 1 & 0 & 0 & 0 \\ 0 & 1 & 0 & 0 \\ 0 & 0 & 1 & 0 \\ 0 & 0 & 0 & 1 \end{bmatrix} \end{aligned}$$

in which the number of permutation matrices $K = 4$ is reduced. Again, applying the *HuRR* scheduling, the resulting sequence is

$$P_1 P_4 P_3 P_4 P_2 P_4 P_3 P_4$$

which gives rise to the following two-dimensional scheduled tokens

$$\begin{array}{cccccccc} b & a & a & a & c & a & a & a \\ c & b & d & b & d & b & d & b \\ a & c & b & c & b & c & b & c \\ d & d & c & d & a & d & c & d \end{array}$$

The smoothness measures of this two-dimensional scheduling are given by

$$\overline{D} = \left[ \begin{array}{cccc} 1.1250 & 1.4375 & 1.7500 & 1.0267 \end{array} \right]^T$$

and

$$\underline{D} = \left[ \begin{array}{cccc} 1.125 & 1.4375 & 1.4375 & 1.3392 \end{array} \right]$$

which sum up to the overall smoothness $D = 5.3392$. By comparing the output smoothness $\underline{D}$ with that of the previous decomposition, we find some improvement in the fourth output. Similar improvement can be found at input module 3, where the optimal smoothness $\overline{D}_3 = 1.7500 = \overline{H}_3$ is achieved. The overall smoothness $D = 5.3392$ is also better than the previous $D = 5.3794$ but still greater than the entropy of capacity matrix $H(C) = 5.1714$.

Preceding examples indicate that the performance of path switching depends not only on the scheduling algorithm but also on the number of permutation matrices in the capacity matrix decomposition. As we mentioned above that the number of permutation matrices in the case of uniform capacity matrix with maximal entropy is $K = N$, and the worst delay bound is on the order of $O(K)$ for most known scheduling algorithm, it is therefore reasonable to arrive at the conjecture that $K$ is on the order of $O(N)$ for the optimal decomposition of any capacity matrix, even though the known bound of $K$ in the Birkhoff von-Neumann decomposition is $N^2 - 2N + 2$. The Kraft matrix and the smoothness measure introduced in this section will provide the objective of optimal two-dimensional scheduling to be explored in the future, which is expected to be a very hard problem because of the large number of possible combinations.

## VII. Conclusion

The parallels between packet switch subject to the contention and transmission channel in the presence of noise are consequences of the law of probability. Input signal to transmission is a function of time, the main theorem on noisy channel coding being based on the law of large number. On the other hand, the input signal to switch is a function of space, and both theorems on deflection routing and smoothness of scheduling are proved on the ground of randomness. In both systems, the random disturbances are tamed by means of similar mathematical tools aimed for reliable communication.

The communication network has gone through two phases of quantization in the last century. The first phase was the quantization of transmission channels, from analog to digital based on the sampling theorem of bandlimited signal. The second phase is the quantization of the switching system, from circuit switching to packet switching. The comparisons provided in this paper demonstrate that the scheduling based on capacity matrix decomposition serves the same function as the sampling theorem to reduce the complexity of communication.

The concept of path switching can further be extended to the network level to cope with resources contentions. A connection-oriented sub-network with predetermined topology and bandwidth can be embedded in the current IP network for supporting QoS of real-time services such as voice or video over IP. The scheduling of path switching in conjunction with the routing scheme of multiprotocol label switching (MPLS) will provide a platform to support different traffic classes of differentiated services (DiffServ) [54], [55]. The predetermined paths of label switched sub-network are similar to the subway system embedded in the public transportation network. It will provide a coherent end-to-end QoS solution to the multilevel resource allocation [14] issues arising from real-time services. First, the predetermined connection patterns at each router are contention-free at the packet level. Next, the periodic connection patterns are frame based, the bandwidth assigned to each session is fixed at burst level within each frame. Finally, the stable capacity of each session is guaranteed at call level. The traffic engineering of this internet subway system could be a challenge networking research area in the future.

## Acknowledgement

The development of this work was helped by discussions over the years with my students Cheuk H. Lam, Philip To, Man

Chi Chan, S. Y. Liew, Yun Deng, Manting Choy, Jianming Liu, Sichao Ruan, and Li Pu, all members of Broadband Communication Laboratory at the Chinese University of Hong Kong.

## APPENDIX I
## BOLZMANN MODEL OF PACKET DISTRIBUTION

Consider an $N \times N$ switch as a thermal system. The outputs are particles and the energy level of an output, denoted by $\varepsilon_i$, equals to the number of packets destined for that output. The distribution of packets over outputs can be determined by maximizing the Boltzmann entropy [24].

Suppose there are $M$ packets sending from $N$ input ports to $N$ output ports. Let $n_i$ be the number of output ports with energy level $\varepsilon_i$. An example with $N = 8$ and $M = 4$ is shown in Fig. 25, where $n_0 = 5, n_1 = 2$, and $n_3 = 1$. The number of possible divisions of $N$ outputs into $r$ distinct energy levels of respective sizes $n_0, \cdots, n_r$ is

$$\frac{N!}{n_0! n_1! \cdots n_r!},$$

and the number of possible divisions of $M$ packets into $N$ distinct outputs of respective sizes $\underbrace{0, \cdots, 0}_{n_0}, \cdots, \underbrace{r, \cdots, r}_{n_r}$ is

$$\frac{M!}{(0!)^{n_0}(1!)^{n_1} \cdots (r!)^{n_r}}$$

Suppose each input can send one packet at most in any time slot, then the total number of states of the entire switch is given by

$$W = \frac{N!}{(N-M)!M!} \cdot \frac{N!}{n_0! n_1! \cdots n_r!} \cdot \frac{M!}{(0!)^{n_0}(1!)^{n_1} \cdots (r!)^{n_r}} \tag{65}$$

subject to the following constraints:

$$N = n_0 + n_1 + n_2 \cdots + n_r, \tag{66}$$

and

$$M = 0n_0 + 1n_1 + 2n_2 \cdots + rn_r. \tag{67}$$

The Boltzmann entropy $S$ of the system is given by:

$$S = C \ln W \tag{68}$$

where C is the Boltzmann constant. The maximal entropy can be obtained from the following function formed by Lagrange multipliers

$$f(n_i) = \ln W + \alpha(\sum_i n_i - N) + \beta(\sum_i i n_i - M). \tag{69}$$

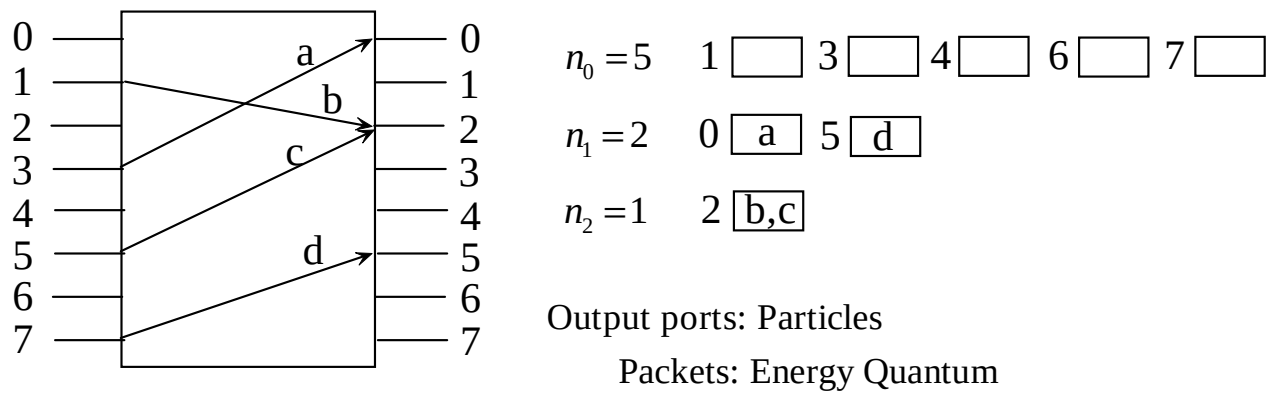

Fig. 25. Blotzmann model of packet distribution

Using Stirling's approximation $\ln x! \approx x \ln x - x$ for the factorials, we have

$$\begin{aligned} f(n_i) \doteq& N \ln N - N - (N-M)\ln(N-M) + (N-M) \\ &+ N \ln N - N - \sum_i (n_i \ln n_i - n_i) - \sum_i n_i \ln(i!) \\ &+ \alpha(\sum_i n_i - N) + \beta(\sum_i i n_i - M). \end{aligned} \tag{70}$$

Taking the derivatives with respect to $n_i$, and setting the result to zero and solving for $n_i$ yields the population numbers:

$$n_i = \frac{e^{(\alpha+\beta i)}}{i!}. \tag{71}$$

If the offered load $\rho$ is uniform on each input, then we have

$$\rho = \frac{M}{N} = \frac{\sum_i i n_i}{\sum_i n_i} = e^{\beta}. \tag{72}$$

The probability that there are $i$ packets destined for a particular output has the Poisson distribution:

$$p_i = \frac{n_i}{N} = \frac{\frac{e^{(\alpha+\beta i)}}{i!}}{\sum_i n_i} = e^{-\rho}\frac{\rho^i}{i!}, \qquad i = 1, 2, 3, ... \tag{73}$$

The carried load equals the probability that an output port is busy

$$\rho' = 1 - p_0 = 1 - e^{-\rho} \tag{74}$$

which is consistent with (1). Hence, our proposition on the signal power of switch is verified.

In the above derivation, the assumption that each input can only send at most one packet in any time slot can be relaxed. As long as packets are distinguishable, the total number of states of outputs is

$$W = \frac{N!}{n_0! n_1! \cdots n_r!} \cdot \frac{M!}{(0!)^{n_0}(1!)^{n_1} \cdots (r!)^{n_r}} \tag{75}$$

and the distribution of packets is still Poisson. On the other hand, suppose all packets are indistinguishable, then the total number of states of outputs becomes

$$W = \frac{N!}{n_0! n_1! n_2! \cdots n_r!}. \tag{76}$$

Following the same procedure, we obtain the geometric distribution of packets

$$p_i = \frac{1}{1+\rho} \cdot (\frac{\rho}{1+\rho})^i, \qquad i = 0, 1, 2, ... \tag{77}$$

which will be inconsistent with (1).

We next show that both the signal $S_p$ and noise $N_p$ of an $N \times N$ crossbar switch are normally distributed under the assumptions of homogeneous input loading $\rho$ and uniform output address.

The signal power $S_p$ is the sum of the following i.i.d. random variables:

$$X_i = \begin{cases} 1 & \text{if output } i \text{ is busy} \\ 0 & \text{otherwise} \end{cases}$$

with mean $E[X_i] = \rho'$ and variance $Var[X_i] = \rho'(1-\rho')$, where $\rho'$ is the carried load given in (1). It follows from central limit theorem that the signal power $S_p = \sum_{i=1}^{n} X_i$ is normally distributed with mean $E[S_p] = N\rho'$ and variance $Var[S_p] = N\rho'(1-\rho')$. The noise power $N_p = N - S_P$ becomes independent of the signal power $S_p$ when $N$ is large. By using the similar argument, the noise $N_P$ is also normally distributed with mean $E[N_p] = N(1-\rho')$, and variance $Var[N_p] = N\rho'(1-\rho')$.